\documentclass[twocolumn]{aastex631}

\newcommand\ec{$\eta$~Car}
\newcommand\hst{{\it HST}}
\newcommand\stis{{\it STIS}}
\newcommand\cmfgen{{\it CMFGEN}}
\newcommand\kms{km~s$^{-1}$}

\newcommand\Vinf{V$_{\infty}$}

\newcommand{\Lsun}{\hbox{$L_\odot$}}
\newcommand{\Msun}{\hbox{$M_\odot$}}
\newcommand{\Mdot}{\hbox{$\dot M$}}
\newcommand{\Msunyr}{\hbox{$M_\odot\,$yr$^{-1}$}}
\received{tbd}
\revised{tbd}
\accepted{tbd}
\submitjournal{ApJ}
\shortauthors{Gull et al.}

\begin{document}
\title{Eta Carinae: the dissipating occulter is an extended structure}
\correspondingauthor{Theodore R. Gull}\email{tedgull@gmail.com}

\author[0000-0002-6851-5380]{Theodore R. Gull}
\affiliation{Exoplanets \&\ Stellar Astrophysics Laboratory, NASA/Goddard Space Flight Center, Greenbelt, MD 20771, USA}\affiliation{Space Telescope Science Institute, 3700 San Martin Drive, Baltimore, MD 21218, US}
\author[0000-0001-9853-2555]{Henrik Hartman}
\affiliation{Materials Science \&\ Applied Mathematics, Malm\"{o}
University, SE-20506 Malm\"{o}, Sweden}
\author[0000-0002-8289-3660]{Mairan Teodoro}\affiliation{Space Telescope Science Institute, 3700 San Martin Drive, Baltimore, MD 21218, US}
\author[0000-0001-5094-8017]{D. John Hillier}
\affiliation{Department of Physics \&\ Astronomy \& Pittsburgh Particle Physics,
    Astrophysics, \&\ Cosmology Center (PITT PACC),\hfill\\  University of Pittsburgh,  3941 O'Hara Street, Pittsburgh, PA 15260, USA}

\author[0000-0002-7762-3172]{Michael F. Corcoran}
\affiliation{CRESST \&\ X-ray Astrophysics Laboratory, NASA/Goddard Space Flight Center, Greenbelt, MD 20771, USA}
\affiliation{The Catholic University of America, 620 Michigan Ave., N.E. Washington, DC 20064, USA}
\author[0000-0002-7978-2994]{Augusto Damineli}
\affiliation{Universidade de S$\tilde{a}$o Paulo, IAG, Rua do Mat$\tilde{a}$o 1226, Cidade Universit$\acute{a}$ria S$\tilde{a}$o Paulo-SP, 05508-090, Brasil}
\author[0000-0001-7515-2779]{Kenji Hamaguchi}
\affiliation{CRESST \&\ X-ray Astrophysics Laboratory, NASA/Goddard Space Flight Center, Greenbelt, MD 20771, USA}
\affiliation{Department of Physics, University of Maryland Baltimore County, 1000 Hilltop Circle, Baltimore, MD 21250, USA}
\author[0000-0001-7697-2955]{Thomas Madura}\affiliation{Department of Physics \&\ Astronomy, San Jose State University, One Washington Square, San Jose, CA 95192, USA}
\author[0000-0002-4333-9755]{Anthony F. J. Moffat}\affiliation{D$\acute{e}$pt. de physique, Univ. de Montr$\acute{e}$al, C.P. 6128, Succ. C-V, Montr$\acute{e}$al, QC H3C 3J7, Canada \& Centre de Recherche en Astrophysique du Qu$\acute{e}$bec, Canada}
\author[0000-0002-5186-4381]{Patrick Morris}\affiliation{California Institute of Technology, IPAC, M/C 100-22, Pasadena, CA 91125, USA}
{\author[0000-0003-2636-7663]{Krister Nielsen}
\affiliation{The Catholic University of America, 620 Michigan Ave., N.E. Washington, DC 20064, USA}}
\author[0000-0002-2806-9339]{Noel D. Richardson}
\affiliation{Department of Physics \&\ Astronomy, Embry-Riddle Aeronautical University, 3700 Willow Creek Rd, Prescott, AZ 86301, USA}
\author[0000-0001-7673-4340]{Ian R. Stevens}\affiliation{School of Physics \& Astronomy, University of Birmingham, Birmingham, B15 2TT, UK}
\author[0000-0001-9754-2233]{Gerd Weigelt}
\affiliation{Max Planck Institute for Radio Astronomy, Auf dem H\"ugel 69, 53121 Bonn, Germany}

\begin{abstract}
Previous STIS long-slit observations of \ec\  identified numerous absorption features in both the stellar spectrum, and in the adjacent  nebular spectra, along our line-of-sight. The absorption features became temporarily stronger when the ionizing FUV radiation field was reduced by the periastron passage of the secondary star. Subsequently, dissipation of a dusty structure in our LOS has led to a long-term  increase in the apparent magnitude of \ec, an increase in the ionizing UV radiation,  and the disappearance of absorptions from multiple velocity-separated shells extending across the foreground Homunculus lobe. We use \hst/STIS spectro-images, coupled with published infrared and radio observations, to locate this intervening dusty structure. Velocity and spatial information indicate the occulter is  $\approx$1000 au in front of \ec. The Homunculus is a transient structure composed of dusty, partially-ionized ejecta that eventually will disappear due to the relentless rain of ionizing radiation and wind from the current binary system along with dissipation and mixing with the ISM. This evolving complex continues to provide an astrophysical laboratory that changes on human timescales.
\end{abstract}


\section{Introduction\label{sec:intro}}
Eta Carinae (\ec) caught the attention of southern observers in the early nineteenth century as its visual magnitude changed frequently \citep{SmithFrew11}. By the 1840s its visual magnitude rivaled Sirius only to disappear to the naked eye, then marginally brightened in the 1890s and faded again. In the period 1939-1945 there was a  jump by 1.2 mag in the visual brightness without brightening of the central star \citep{Thackeray53}. After that the brightening  progressed at a slower pace \citep{Gaviola50} to the present day when it has again achieved  visual magnitude $\approx\ $4.4 \citep{Damineli21}.

The changing properties of \ec\  continue to fascinate astronomers. Since \ec\ is located relatively nearby (2300 pc: \cite{Smith06}), ejecta are resolvable from \ec\ enabling studies not only of the central source, but the influence on material thrown out during historical times. Southern hemisphere telescopes  and space observatories with increasingly sophisticated instruments often turn to this amazing object to gain new information as the spectra of  the central source and its ejecta evolve. The apparent changes of \ec\ and its expanding ejecta make it an astrophysical laboratory that evolves on human timescales enabling new studies of topics ranging from  atomic and molecular physics to the evolution of  massive stars nearing their end stages.

\ec\ ejected at least two shells of material within the historical record. Photographic imagery in the 1940s revealed a surrounding extended structure that \citet{Gaviola50} named the 'Homunculus' as it appeared to be humanoid in shape. Near diffraction-limited imagery of the {\it Hubble Space Telescope} (\hst) in the mid-1990s clarified the Homunculus to be  a dusty,  bipolar 10\arcsec$\times$20\arcsec\ structure with an  equatorial skirt \citep{Morse98}. Spectro-imagery from the \hst/Space Telescope Imaging Spectograph (\stis) identified an internal structure, the Little Homunculus, as a co-aligned, ionized, bipolar 4\arcsec$\times$4\arcsec\ structure  \citep{Ishibashi03}. The Homunculus, by velocity and proper motion, is associated with the Great Eruption of the 1840s \citep{Smith17} and the Little Homunculus is associated with the Lesser Eruption of the 1890s \citep{Ishibashi03}.

\ec\ is a massive binary. The primary component is the closest example of a very massive star. \cite{Damineli96} and \cite{Damineli97} used spectroscopic observations,  correlated with the photometric near-infrared (NIR) light curve,  to discover the 5.5 yr binary cycle of \ec. \cite{Corcoran05a}  confirmed the binary period in X-rays using the Rossi X-ray Timing Explorer (RTXE). RXTE spectra analyzed by \cite{Ishibashi99a} showed that the wind velocity of the companion star was at least a factor of 3 higher than the wind velocity  of the primary star, while comparison of Chandra X-ray grating spectra with wind-wind collision models by 
\citet{PIttard02} provided mass loss rate estimates, \Mdot$_A \approx\ $2.5 $\times$ 10$^{-4}$ \Msun\ and \Mdot$_B \approx$  10$^{-5}$ \Msun\  yr$^{-1}$ with \Vinf$_{,A}\ \sim$ 500 to 700 \kms and \Vinf$_{,B}\ \sim$ 3000 \kms. 

More recent refinements are: the binary period, P = 5.53 yr \citep{Teodoro16} with  the primary mass $\gtrsim$ 100 \Msun\ \citep{Hillier01, Hillier06}, \Mdot$_A =~$8.5 $\times$ 10$^{-4}$ \Msunyr\ \citep{Madura13}  and \Vinf$_{,A}$ = 420 \kms\ \citep{Groh12a}. The secondary, \ec-B, is estimated to have a mass $\approx$ 30 to 50 \Msun\ \citep{Verner05a, Mehner10}. The secondary properties remain inferred as direct measure of the secondary is yet to be accomplished because of the huge contrast in brightness between the companions. A possible exception may be the recent study by \cite{Strawn23} that tracked the weak \ion{He}{2} $\lambda$4686 emission to directly measure the radial velocity orbit of the secondary, which moves in anti-phase relative to the primary star: they find a minimum mass of 55 \Msun\ for \ec-B with a minimum mass for \ec-A of 102 \Msun.

The observed optical and UV flux of \ec,  monitored by multiple facilities from the ground and in space, has increased over the past two decades. The response of the dusty Homunculus, serving as a calorimeter to the 6~$\times\ $~10$^6$ \Lsun\ source, has been constant for nearly a half century  other than a possible 5.54-yr modulation due to the binary orbit \citep{Mehner19}.  The X-ray light curve also has been repeatable \citep{Corcoran10, Espinoza21a}. \citet{Damineli19} noted that while the flux of \ec\ at visible wavelengths increased by an order of magnitude, the contribution of scattered starlight from the Homunculus stayed constant, demonstrating that the recent changes in \ec\ were apparently due to a dusty occulter that has  been dissipating and/or moving out of the line of sight (LOS). 

\cite{Hillier92a}, \cite{Hillier01} and \cite{Hillier06} previously noted that the equivalent width of H$\alpha$\ measured directly of \ec\ was inconsistent with atmospheric models by \cmfgen\ and was larger than  measured in the scattered starlight off the Homunculus. They suggested that an occulting structure, with very grey spectral properties, systematically blocked the source of the stellar continuum more than the very extended wind of \ec-A. 

Furthermore, \cite{Weigelt95} suggested the existence of an occulting structure in front of Eta Car to explain the unusually high relative brightness of three ejected clumps at separations of 0\farcs1 to 0\farcs3 \citep{Weigelt86}.  These three clumps were only 1.3 to 2.4 mag fainter than the central star Eta Car in the wavelength region of 8000 to 9000\AA\ in 1983 when they were discovered  and only a factor of two fainter than Eta Car in the UV at ~2000\AA\ in 1991 and 1992 \citep{Davidson95a}.

The long-term decrease of the emission lines and the increase in optical  depth of P Cygni absorption were replicated by a model simulating a dissipating occulter in front of a stable primary star \citep{Damineli22}. \cite{Damineli21} predicted that by the 2030s, the occulter will have completely dissipated. Recently \cite{Pickett22} showed that a long-term evolution of the Na D absorption from the Little Homunculus was consistent with a disappearing occulter.
 
The tenfold increase in FUV\footnote {Following the \hst/STIS echelle spectral modes, we  define NUV to be 1750 to 3150\AA\ (E230H, E230M), FUV to be 1215 to 1750 \AA\ (E140H, E140M) and the EUV to be shortward of 1216 \AA\ (Ly $\alpha$).} \citep{Gull21a,Gull22} since 2000 has affected many of the three dozen shells  ranging in velocity from $-$121 to $-$1074 \kms\ in the LOS as catalogued in the NUV by \cite{Gull06}.  Nielsen et al. (in prep.) found that nearly all of the $\sim$800 strong absorption lines of H$_2$ ($-$513 \kms, associated with the Homunculus)  disappeared and most low-ionization  signatures of absorption systems between $-$122 and $-$186 \kms,  associated with the Little Homunculus,   likewise disappeared. Current evidence suggests that slower-moving shells, being closer to \ec, are now more highly-ionized and that the H$_2$ has been destroyed by the increased EUV and FUV radiation escaping from the central region. 

In this discussion we examine archived NUV spatially-resolved \hst/STIS spectra (spectro-images)\footnote{Based on archived observations made with the NASA/ESA Hubble Space Telescope, obtained at the Space Telescope Science Institute, which is operated by the Association of Universities for Research in Astronomy, Inc., under NASA contract NAS5-26555.}  that include portions of the foreground ejecta to estimate the spatial extent of the dissipating occulter. Section \ref{sec:geometry} reviews the geometry of the Homunculus and Little Homunculus. Section \ref{sec:obs} catalogues the selected spectro-images that provide temporal information on the dissipating occulter. Section \ref{sec:longslit} describes spatially-resolved, high-velocity nebular structures  that respond to the high- and low-ionization states of the 5.5-year binary period and that exhibited long-term changes. 
Section \ref{sec:LOS} provides insight on where the occulter/absorber is located.   A brief conclusion is  in Section \ref{sec:SUM}.

\section{Setting the stage}{\label{sec:geometry}}
\begin{figure}[ht]
\includegraphics[width=8.5cm]{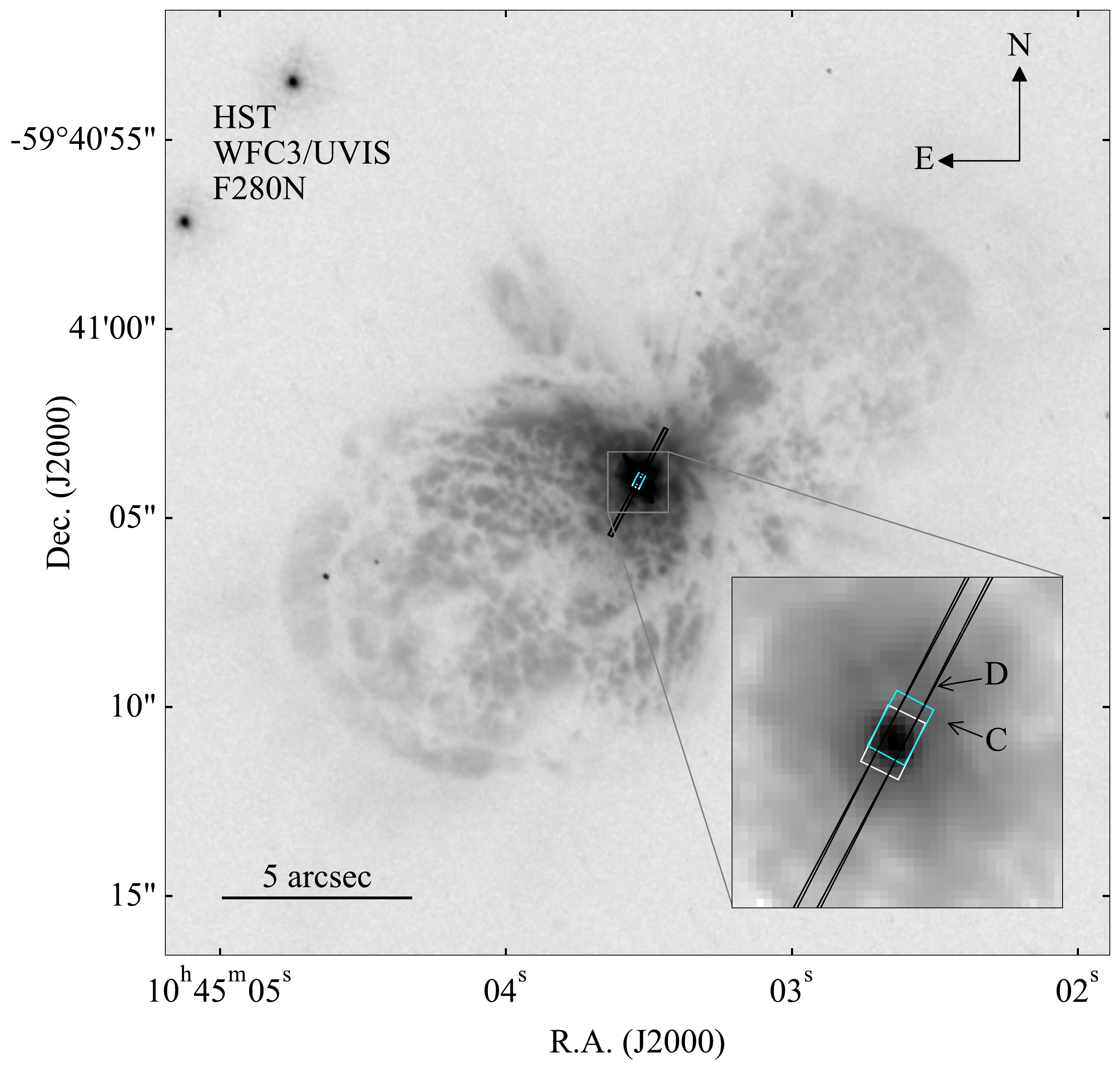}
\caption{ HST image of the Homunculus and the central source recorded with the WFC3/UVIS instrument and the F280N filter on 2020-02-12 ($\Phi=13.997$). The 3\farcs2-long portion of the 52\arcsec$\times$0\farcs1 aperture is indicated by the long, black rectangle (used for the spectro-images displayed in Figures \ref{fig:hilo}, \ref{fig:long}, \ref{fig:occult}, A.1, and A.2). The shorter white  and cyan rectangles (see enlargement) indicate the positions  of the 0\farcs3$\times$0\farcs2 aperture used with the \hst/STIS echelle for the spectro-images displayed in Figure \ref{fig:occult}, top and bottom. The position of Weigelt C and D are indicated in the 1.6\arcsec$\times$1.6\arcsec inset image.   The projected edge of the foreground lobe is located $\approx$ 1\farcs3 NNW of \ec. (Image is from \hst\ archive.)
\label{fig:image}}
\end{figure}

 \begin{figure*}[ht]
\includegraphics[width=18cm]{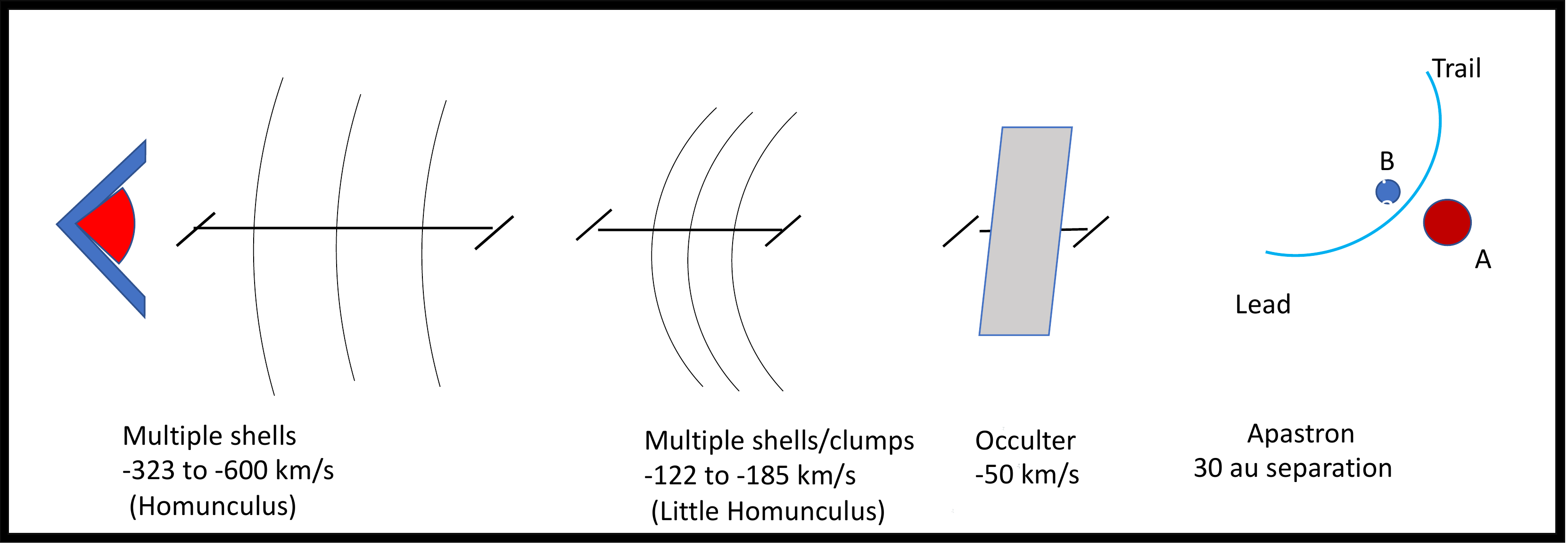}
\caption{ 
Sketch of structures within the Homunculus as seen in our LOS ({\bf NOT TO SCALE}). Is the occulter a portion of the Little Homunculus or a dense clump, like the Weigelt clumps that appear to be within several hundred au of the binary?  In NUV spectra recorded between 2000 and 2004, multiple, weaker shells at various velocities clustered near $-$513 \kms, associated with the Homunculus and the 1840s Great Eruption, were identified \citep{Gull06}. Additional shells clustered  in velocities above and below $-$146 \kms\ are associated with the Little Homunculus ejected in the 1890s Lesser Eruption. By 2018, signatures of most of the weaker shells had disappeared. For scale, the strongest component of the Homunculus,  moving at $-$513 \kms\ since the 1840s, is about 1.8 $\times$ 10$^{4}$ au and the  Little Homunculuus, moving at $-$146 \kms\ since the 1890s, is about 3 $\times$ 10$^3$ au in front of  \ec. The occulter and the interacting winds are associable with or within the Little Homunculus (see Figure \ref{fig:vellos}).  The reader will note that the shape and extent of the occulter remains uncertain.}
\label{fig:sketch}
\end{figure*}
 We view the bipolar Homunculus with its axis of symmetry projecting at $\approx45\degr$ to the sky plane and tilted on the sky with the foreground lobe in the southeast. \ec\ is physically centered between the two lobes with its orbital plane perpendicular to the Homunculus axis of symmetry \citep{Madura12}. From our vantage, the stellar position projects, as shown in Figure \ref{fig:image}, close to an edge of the foreground lobe, leading to a noticeable velocity shift across the nebular structure as shown below. Immediately surrounding \ec\ is a complex of dusty ejecta that strongly scatters starlight out to 1\arcsec\ diameter (2300 au). Clumps of emission are scattered throughout this complex, the brightest of which are Weigelt C and D, two   clumps first noticed by near-red speckle interferometry to be nearly as bright as \ec\ itself \citep{Weigelt86}. Proper motion studies, most recently done by \cite{Weigelt12} indicate these clumps were ejected in the Lesser Eruption. \cite{Davidson95a}, using data from the \hst/Faint Object Spectrograph (FOS), determined these to be emission-line structures photo-ionized by \ec. Detailed mapping of the emission clumps in [\ion{Fe}{2}] and [\ion{Fe}{3}] and the variation in ionization across one orbital period were described by \cite{Gull16}.
 
 The long, thin rectangle, superimposed on Figure \ref{fig:image}, defines the position of a 3$\farcs2$-long segment of the \hst/STIS 52$\arcsec\times$0\farcs1 aperture oriented at PA $=~$332\degr, marking the portion of the Homunculus sampled in the spectro-images displayed in Figures \ref{fig:hilo}, \ref{fig:long}, \ref{fig:apcompare} and \ref{fig:ap2}. The aperture extends from south by southeast (SSE) across the position of \ec\ to north by northwest (NNW) ending at the apparent edge of the foreground lobe of the Homunculus. This aperture position had been selected to monitor changes in Weigelt B and D, located 0$\farcs$2 and 0$\farcs$3 respectively to the NNE of \ec, as their ionization changed between  the high-ionization and low-ionization states by direct radiation from  \ec-B. The Weigelt B clump is no longer detected.

 The white rectangle in the enlargement within Figure \ref{fig:image} outlines the position of the 0\farcs2$\times$0\farcs3 aperture centered on \ec\ for the \hst/STIS echelle spectro-image recorded at $\phi  =$ 11.037, a portion of which is displayed in Figure \ref{fig:occult}, top. The cyan rectangle shows the position for the spectro-image recorded at $\phi =$ 11.119 (Figure \ref{fig:occult}, bottom),  offset 0\farcs1 NNW from \ec, which placed the star at the edge of the aperture and allowed sampling  the changes in front of Weigelt  B and D during the early recovery to high-ionization state.
 
A sketch, displayed in Figure \ref{fig:sketch}, relates the  locations in the LOS from \ec\ of the two major   shell complexes, the Homunculus ($-$513 \kms) and the Little Homunculus ($-$146 \kms). \cite{Gull06} found many fainter  velocity components in absorption lines of singly-ionized metals clustering around these two strongly-defined velocities  in NUV (\hst/STIS  E230H spectra) recorded between 2000 and 2004. The dissipating occulter, of unknown geometry sketched as a grey rectangle, led to weakening or disappearance of  absorptions at all velocities. In FUV spectra (\hst/STIS G230MB) recorded from 2018 to 2019, the weakened $-$513 \kms\ and very weak $-$146 \kms\ velocity components remain. 

\section {Observations}\label{sec:obs}

\begin{table*}
\tablenum{1}
\centering
\caption{Log of \hst/\stis\ observations used in this study\label{tab:obs}}
\begin{tabular}{lllllrc}
\hline
 Date &HST Program&$\phi$ $^{a}$
 & Aperture&Gratings&PA$^{b}$&Central Wavelength\\\hline
\hline
2000-03-20&8327&10.404 & 52\arcsec$\times$0\farcs1&G230MB&$+$332\degr&2557\AA, 2697\AA\\

2003-09-21&8327&11.037&0\farcs3$\times$0\farcs2&E230H&$+$154\degr&2762\AA\\
2003-09-22 &9973& 11.037 & 52\arcsec$\times$0\farcs1&G230MB&$+$153\degr&2557\AA\\
2004-03-06&9973&11.119&0\farcs2$\times$0\farcs2&E230H&$+$333\degr&2762\AA\\
2018-04-21 &15067& 13.641 & 52$\times$0\farcs1&G230MB&$+$333\degr&2557\AA, 2697\AA\\
2019-06-10 &15611& 13.886 & 0\farcs2$\times$0\farcs2  & E140H & +82\degr & 1234\AA, 1416\AA, 1598\AA\\

\hline

\end{tabular}
\\
$^{a}$ $\phi$ refers to the binary orbital phase plus spectroscopic cycle based upon both the disappearance of He II emission and the X-ray drop with periastron passage numbered 13 occurring on JD 2456874.4 $\pm$ 1.3 days and orbital period of 2022.7 $\pm$ 0.3 days  \citep{Teodoro16}. While in general agreement on the period, observations and models do not concur on the actual periastron event which is thought to be several  days earlier than this reference date. The periastron number refers to the convention established by \cite{Groh04} based upon spectroscopically detected periastrons beginning in February 1948.\\
$^{b}$ The position angle of the aperture (PA). Note that some observations were recorded at or near the complementary PA $=$ 152\degr. Where necessary the spectro-image was mirror-imaged in figures to match spatial position.\\
\end{table*}

This study was motivated by the desire to understand what was changing within the Homunculus, given the noticeable increase in flux in the FUV  \citep{Gull20} and whether archived spectra could provide structure and location of the occulter. Multiple spectra had been recorded of \ec\ and the Homunculus by various programs with the \hst/STIS beginning in early 1998 extending through 2021. We examined each of the spatially-resolved spectra looking for evidence of absorption signatures from the multiple shells catalogued in the LOS by \cite{Gull06} and changes thereof with time. Of the hundreds of \hst/STIS spectra recorded of \ec,   we list in Table \ref{tab:obs} the spectro-images that showed the most informative spatial and/or velocity changes in absorption from high- to low-ionization state and/or across the long term, nearly two decades of time. Spectro-images at the same position angle, or  its 180 \degr\ compliment,  were required. 

The scattered starlight drops an order of magnitude within a few arcseconds of \ec. While in early observations the full spectral range of the CCD was sampled from 1713~\AA\ to 10,000~\AA, the useful spectral absorption signatures of the intervening shells proved to be from singly-ionized metal lines in the 2500 to 2800~\AA\ spectral interval. The strongest, best defined absorptions originated from resonant or near-resonant energy levels. 

The \stis\ 52\arcsec$\times$0\farcs1 aperture sampled extended structure centered on \ec, whose characteristics  change rapidly with position angle. Fortunately a series of observations was systematically centered on \ec\ at position angles near  PA~$=$~332\degr\ or  its complement, $+$152\degr, during the extended high-ionization state and the transient low-ionization state associated with the binary periastron passage. The CCD pixels sampled at 0\farcs05 intervals, which was an adequate match with the 0\farcs1 width of the aperture. The nominal spectral resolving power was R $= \lambda/\delta\lambda\ =~$8,000. Spectro-images recorded along the central portions of the aperture are presented in Figures \ref{fig:hilo}, \ref{fig:long}, \ref{fig:apcompare}  and \ref{fig:ap2}.

Higher spectral and spatial resolutions were obtained with the \stis\ echelle modes at  resolving power, R$~=~$110,000, using the 0\farcs3$\times$0\farcs2 aperture (reductions of which are not supported by STScI), in combination with the E230H grating mode. The NUV Multi Anode Multi Array (MAMA) echelle data were processed by STIS science team software  at a spatial resolution of  0\farcs075 sampled at intervals of 0\farcs0125 along the projected width of the aperture resulting in spatially-resolved imagery  displayed in Figure \ref{fig:occult}. Line-by-line spectral extractions were accomplished using \stis\ science team tools which enabled spatially-resolved spectra of specific structures as imaged through the aperture. 
\section{Absorption changes within the Homunculus and the Little Homunculus\label{sec:longslit}}

In this section, we highlight changes in high velocity absorptions ($>$ 100 \kms) that occur within the  the spatially-extended Homunculus foreground lobe due to FUV-induced ionization. Section \ref{sec:hilo} describes changes brought on by the transient decrease of FUV radiation from the high-ionization state across cycle 10 to the low-ionization state at the beginning of cycle 11 when the FUV radiation from the secondary star is briefly absorbed by the extended primary wind. In Section \ref{sec:long},  we describe the strong decrease and velocity-narrowing of absorptions between the high-ionization states of cycles 10 and 13 triggered by the long-term ten-fold increase in FUV radiation. Much stronger absorption complexes exist in the 2000 to 2200\AA\ spectral region but saturation, along with increased crowding of absorption lines, provided limited information on spatial variations with time.

\begin{table}[ht]
\tablenum{2}
    \centering
    \caption{Identified lines of interest}
 Figure \ref{fig:hilo}  and \ref{fig:long} lines\\
   \begin{tabular}{lrcr}
Ion & Wavelength & Energy& Lower\\
& & Levels & Energy\\
     \hline
Mn II & 2576.88 & 4s a$^7$S$_3$ -- 4p z$^7$P$_4$ & 0 eV\\
Fe II &2756.55 & 4s a$^4$D$_{7/2}$ --  4p z$^4$F$_{9/2}$ & 1.0 eV\\
\hline\\
\end{tabular}
\\
Figure \ref{fig:occult} line\\
\begin{tabular}{lrcr}
     Ion & Wavelength & Energy & Lower\\
     && Levels& Energy\\
\hline
Fe II& 2728.35
   & 4s a$^4$D$_{5/2}$ -- 4p z$^4$D$_{3/2}$ & 1.0 eV\\
\hline\\
\end{tabular}\\
Figure \ref{fig:eV} lines\\
    \begin{tabular}{lrcr}
       Ion & Wavelength & Energy& Lower\\
       &&Levels&Energy\\
\hline
Mg I & 2852.96 & 3s$^2$ $^1$S$_0$ -- 3s3p $^1$P$_1$ & 0.0 eV\\
Fe II& 2756.55 & 4s a$^4$D$_{7/2}$ -- 4p z$^4$F$_{9/2}$ & 1.0 eV\\
Cr II& 2844.08 & 4s a$^6$D$_{7/2}$ -- 4p z$^6$F$_{9/2}$ & 1.5 eV\\
\hline\\
\end{tabular}\\
Figure \ref{fig:vellos} lines\\
\begin{tabular}{lrcr}
       Ion & Wavelength & Energy& Lower\\
       &&Levels&Energy\\
\hline
Si II & 1309.276 & 3s$^2$3p $^2$P$_{3/2}$ -- 3s3p$^2$ $^2$S$_{1/2}$ & 0.04 eV\\ 
C IV & 1550.772 & 2s  $^2$S$_{1/2}$ -- 2p $^2$S$_{1/2}$ & 0 eV\\ 
\hline\\
\end{tabular}
    \caption{Strong absorption lines of interest.  
    \label{tab:lines}}
\end{table}
In the following two subsections, we discuss the multiple changes using two isolated metal absorption lines (Table \ref{tab:lines}).  \begin{itemize}
\item The \ion{Mn}{2} $\lambda$2577 line, which originates from the ground state (0 eV), is  representative of resonant metal lines found in or near the spectral interval from $\lambda\lambda$2570 to 2610 (see  Figure \ref{fig:apcompare} and Table \ref{tab:alines}). 
\item The \ion{Fe}{2} $\lambda$2757 line, which  originates from an energy level 1.0 eV above the ground state, is representative of metal lines found in or near the spectral interval from $\lambda\lambda$2730 to 2760 (see Figure \ref{fig:ap2} and Table \ref{tab:alines}).
\end{itemize}

The behavior of resonant and near-resonant absorption lines, while similar, is sufficiently different to reveal changes in ionization and excitation in the multiple velocity-defined shells originally catalogued by \cite{Gull06}. Of the thirty-six velocity-defined shells identified in the LOS, most velocities clustered around $-$146 \kms\ and $-$513 \kms\ with velocity separations averaging 12 to 15 \kms. The velocity resolution of the longslit spectra using the \stis\ G230MB is about 40 \kms\ in this spectral interval, which is insufficent to resolve individual velocity components. Only the echelle spectro-images, that provided 3 \kms\ resolution in the LOS from \ec, separated the individual components (see Section \ref{sec:echelle}). The CCD spectro-images only prove that the multiple, faint velocity components are present or absent.

\subsection{Absorption changes from  high- to low-ionization state}{\label{sec:hilo}}
\begin{figure*}[ht]
\includegraphics[height=7.0cm]{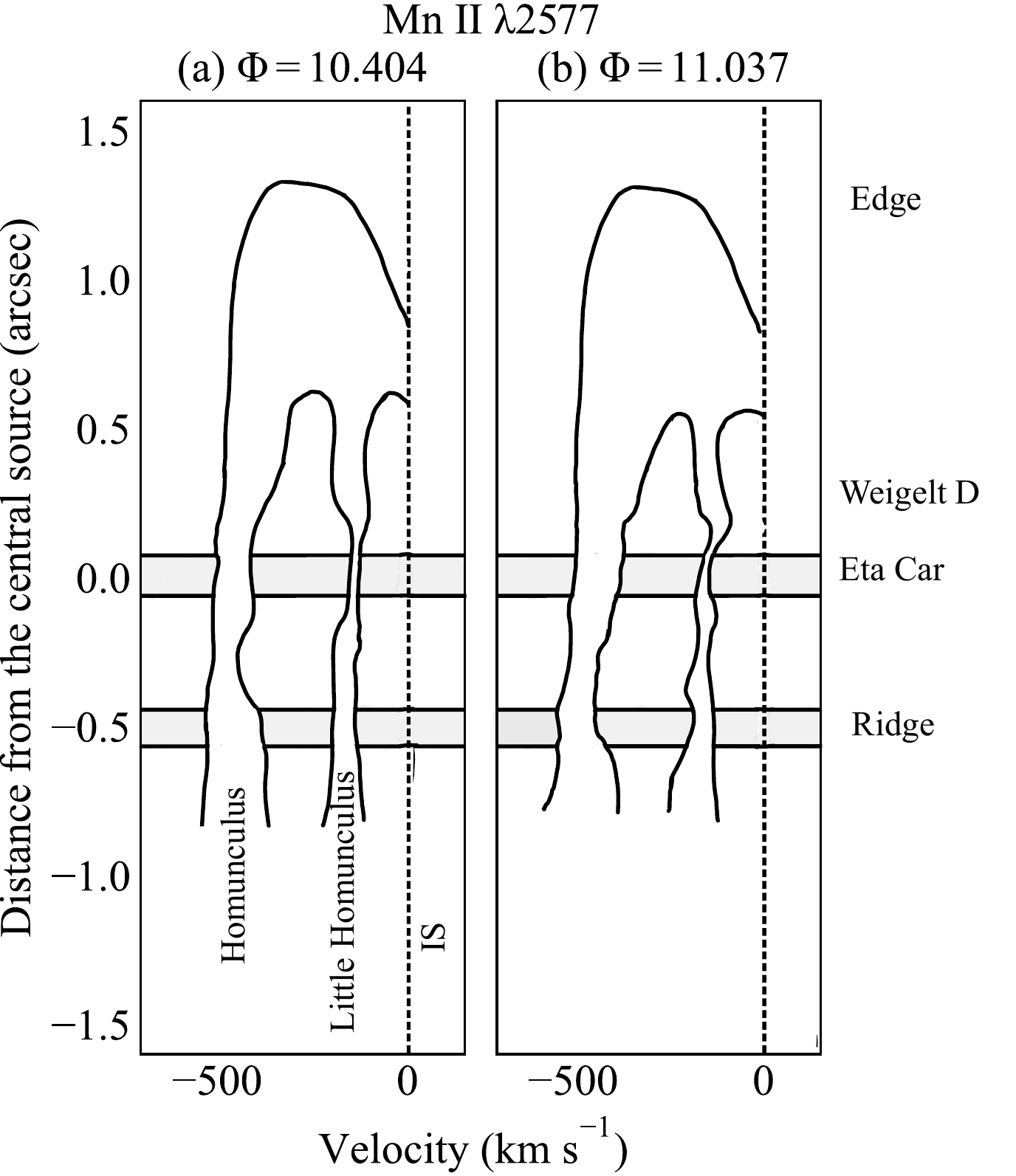} %
\includegraphics[height=7.0cm]{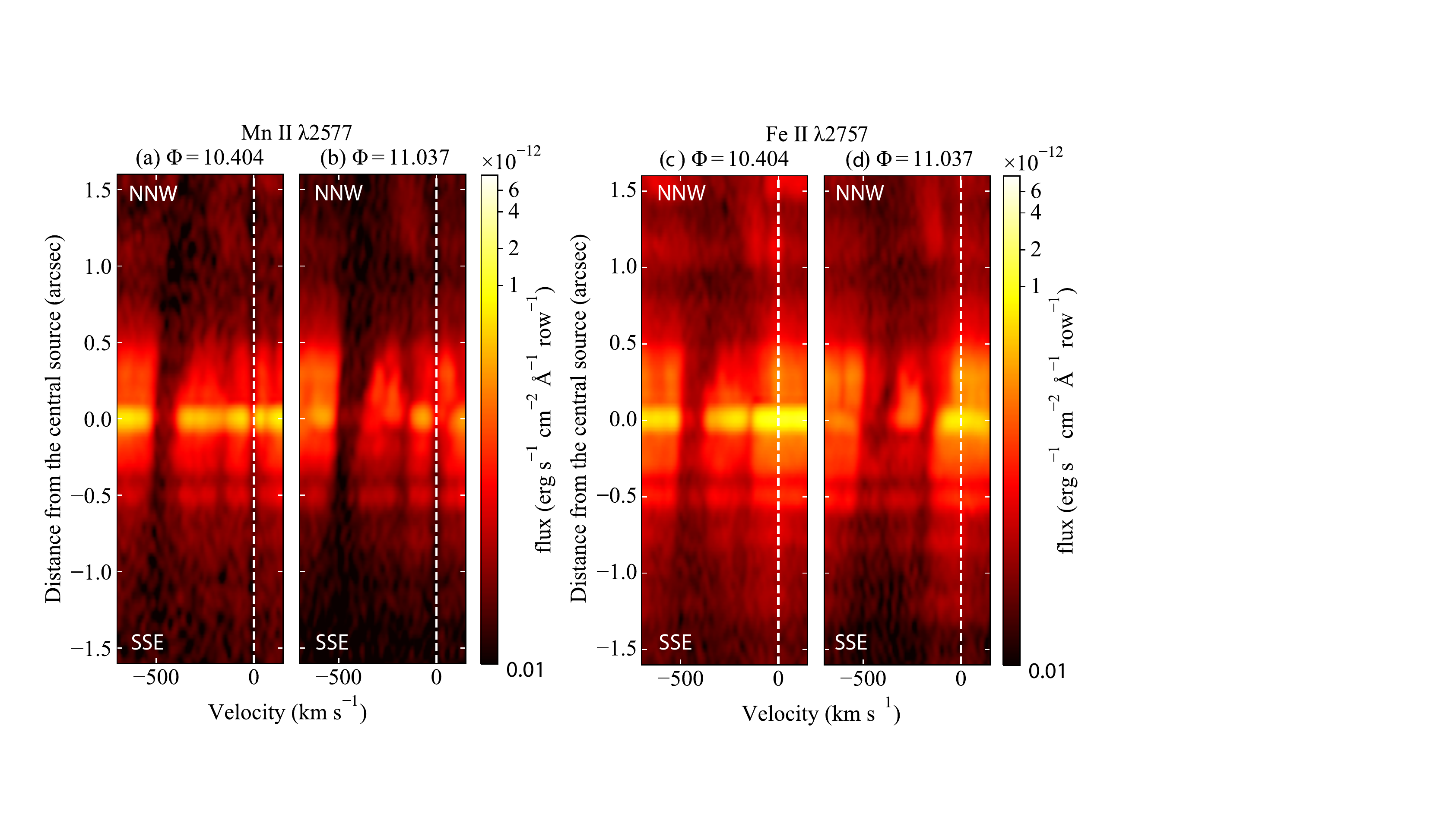}
\caption{Phase-dependent changes from high- to low-ionization state of spatially-varying absorption by foreground ejecta. $\mathbf {Left:}$ Sketch of the absorption structures between the two states. Velocities are referenced to the ISM $\approx\ $ 0 \kms. $\mathbf {Middle:}$ Changes occur in \ion{Mn}{2} $\lambda$2577 line, which originates from the ground state, 0.0 eV, in the short term between (a) the high ionization state at $\phi =$ 10.404 and (b) the low-ionization state at $\phi =$ 11.037. $\mathbf {Right:}$ Changes occur in \ion{Fe}{2} $\lambda$2757 line which originates from a level 1.0 eV above the ground state: (c) $\phi =$ 10.404, (d) $\phi =$ 11.037.  The dashed line in each spectro-image represents 0 \kms.  The sketch includes labeling of the projected Homunculus edge, located $\approx$1\farcs3 NNW of \ec\ and the ridge of backscattered starlight which is discussed in section \ref{sec:back}. The spectro-images at $\phi =$11.037, recorded at PA $=$ 152\degr, are mirror-imaged in this figure for comparison with the spectro-images at $\phi =$ 10.404, recorded at PA=332\degr.}
\label{fig:hilo}
\end{figure*}
Absorptions by singly-ionized metals increase from  high- to low-ionization state due to the drop in EUV and FUV radiation when the EUV-dominating secondary briefly plunges into the extended primary wind.  NUV spectro-images, as displayed in Figure \ref{fig:hilo}, qualitatively show increase in nebular absorption from $\phi =$~10.404, deep in the high-ionization state, to $\phi =$~11.037, deep in the low-ionization state. 

Fluxes in the spectro-images in Figure \ref{fig:hilo} are displayed with a matched log(flux) scale to emphasize the changes in absorption. However the nebular continuum, which apparently scatters from dust on the far side of \ec, is very bright across the central arc-second, then drops by an order of magnitude beyond this region. The edge of the foreground Homunculus lobe ends approximately 1\arcsec\ NNW of \ec. The polar cap is about 4\arcsec\ SE of \ec, while the aperture is oriented from NNW to SSE. 

Most resonant absorption lines include a  component from a thermally cold, foreground cloud near 0 \kms\ which is  marked by the vertical dashed lines in Figure \ref{fig:hilo}. An additional absorption velocity, $+$87 \kms, also is present as identified by \cite{Walborn02} in the Carina Nebula and more recently in Na~D $\lambda$5898 by \cite{Pickett22}. 

A broad absorption feature extends from $\approx-$550 \kms\ at the bottom of each spectro-image (SSE towards the cap of the Homunculus foreground lobe), shifts to $-$500 \kms\ at the position of \ec, broadens while shifting to much lower velocities towards the apparent edge of the foreground lobe ($\approx$ 1\arcsec\ NNW of \ec), then loops towards 0 \kms. A second, broad absorption component begins at the bottom of the spectro-images near $-$160 \kms, shifts slightly to $-$150 \kms\ at the position of \ec\ then shifts toward  0 \kms\  at the edge of the foreground lobe. The strengths of the broad Homunculus $-$500 \kms\ and the Little Homunculus $-$150 \kms\ absorptions increase from the high-ionization state, $\phi =$ 10.404 (Figure \ref{fig:hilo}a) to the low-ionization state, $\phi =$ 11.037 (Figure  \ref{fig:hilo}b). Changes in the absorptions originating from levels near 1 eV are represented by the \ion{Fe}{2} $\lambda$2757 velocity profiles (Figure \ref{fig:hilo}c, d).

\cite{Gull06} measured  kinetic temperatures of the $-$512 \kms\ component of the Homunculus (760 K) and the $-$146 \kms\ component of the Little Homunculus (6370 K) during the cycle 10 high-ionization state. In response to the drop in FUV radiation across the periastron event, the Little Homunculus kinetic temperature dropped to 5000 K.  The kinetic temperature of the Homunculus did not change, but most of the H$_2$ absorptions, driven by FUV excitation, disappeared across the low-ionization state. Note that the lower-velocity absorptions have higher kinetic temperatures (Little Homunculus) than the high higher-velocity absorptions (Homunculus) which is consistent with the faster moving shells being more distant from \ec.

The Homunculus absorption profiles for \ion{Fe}{2} $ \lambda$2757 (Figure \ref{fig:hilo}c, d) are similar to that of  \ion{Mn}{2} $ \lambda$2577 (Figure \ref{fig:hilo}a, b) but  shallower for both the high- and low-ionization states. Much more substructure within individual shells (so called as the structures, somewhat clumpy in nature, appear to define individual surfaces, each with a characteristic velocity within the Homunculus) is noticeable, suggesting that each shell is affected by small variations in velocity, clumpiness and excitation. The low-ionization state, $\phi =$ 11.037 (Figure \ref{fig:hilo}d), shows changes in structure, notably stronger absorptions shift to the red relative to the $-$200 to $-$500 \kms\ absorption just NNW of \ec. The changes in structure appear to be continuous, not abrupt, across the position of \ec, indicating extended structures, most likely shells, not individual clumps.

Changing absorption profiles for additional resonant and near-resonant absorption lines are shown in Appendix \ref{sec:app}.

\subsection{Absorption changes across the long term}\label{sec:long}
\begin{figure*}[ht]
\includegraphics[height=7.0cm]{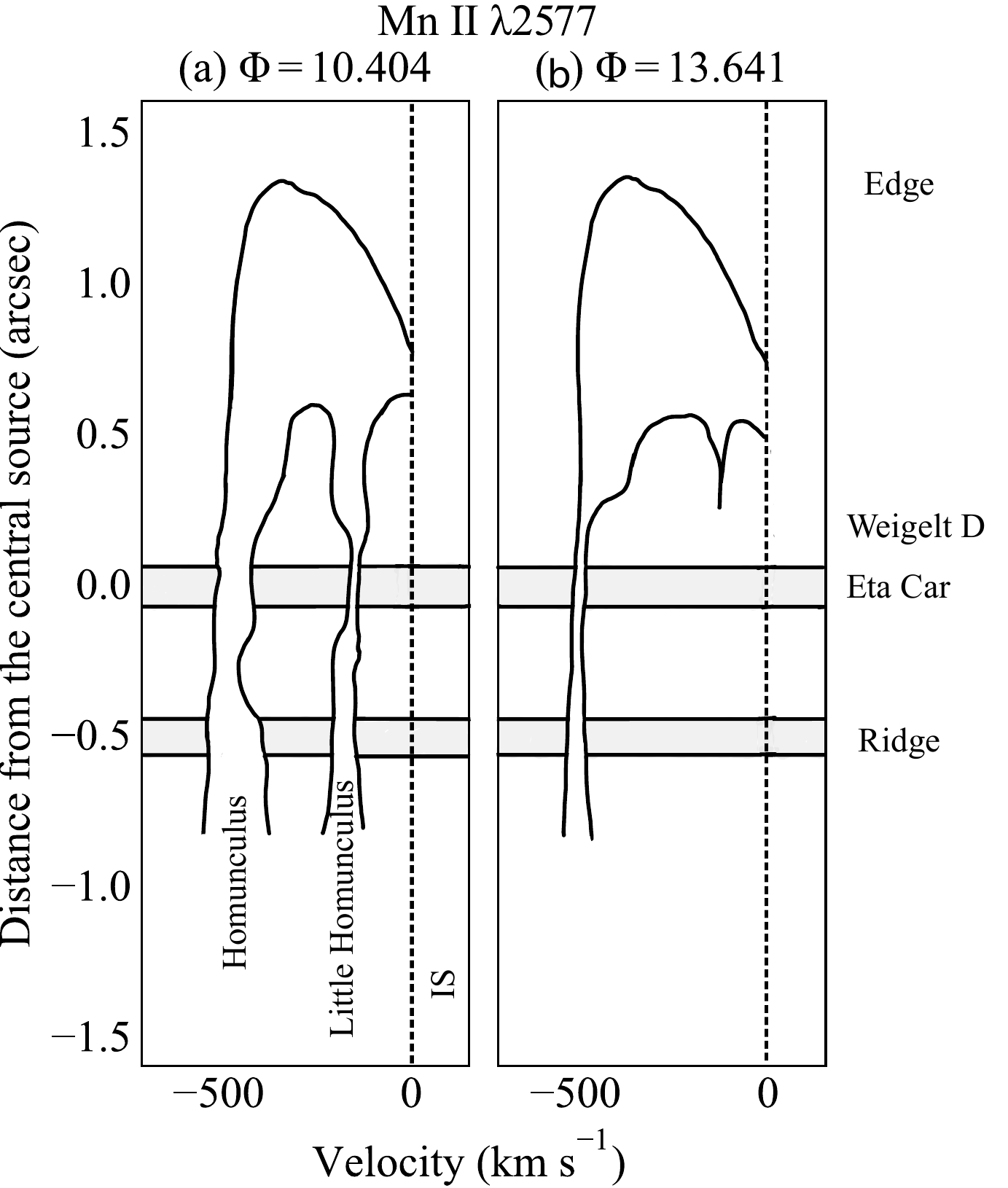} 
\includegraphics[height=7.0cm]{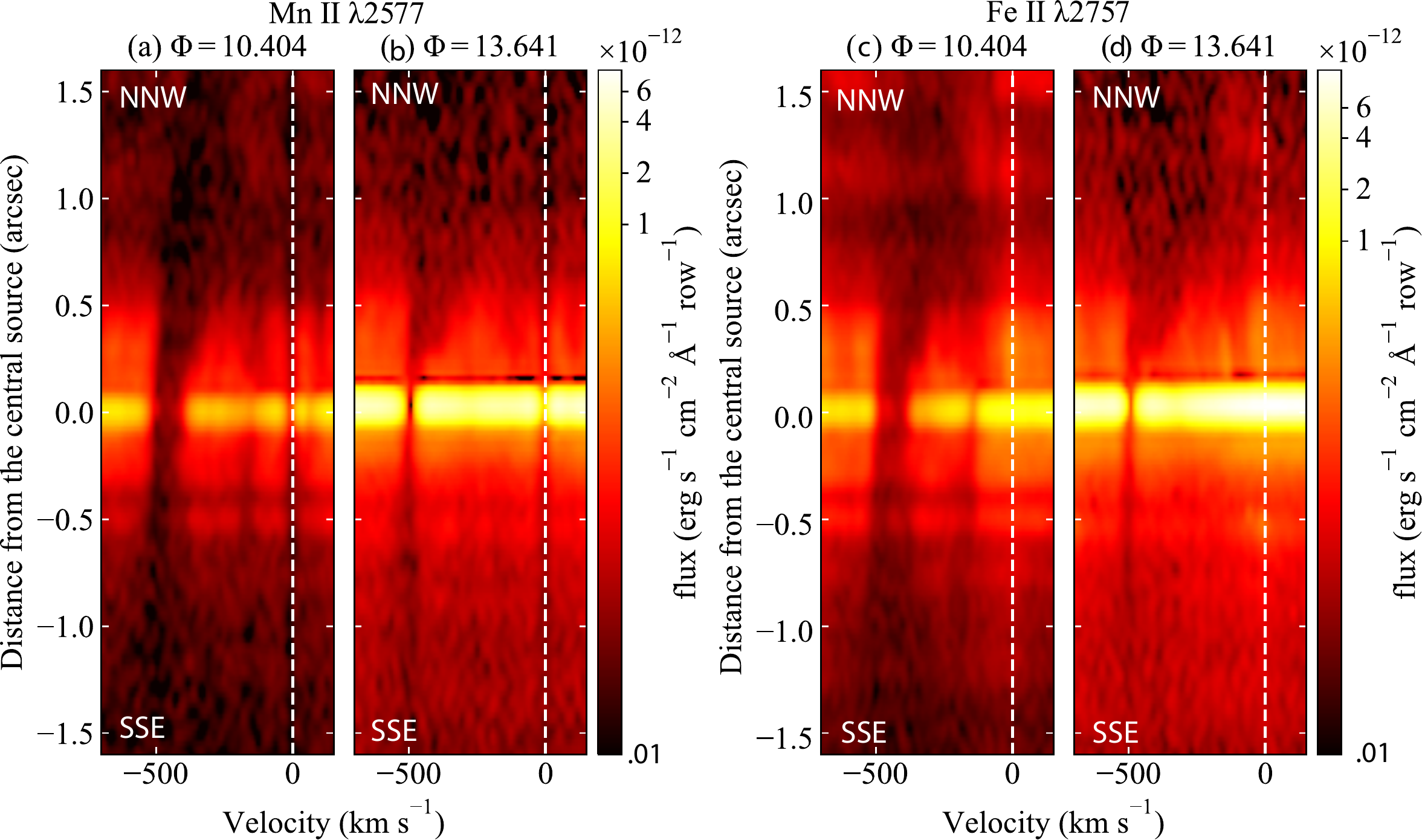}
\caption{Long-term, $\sim$18 years interval, changes of spatially-varying absorption by foreground ejecta between the high-ionization states of Cycles 10 and 13. $\mathbf {Left:}$ Sketch of the absorption structures and how they change over the long term. $\mathbf {Middle:}$ Strong changes occur in the resonant \ion{Mn}{2} $\lambda$2577 line between (a) the high ionization state at $\phi =$ 10.404 and (b) the high ionization state at $\phi =$ 13.641. $\mathbf {Right:}$ Similar changes occur in \ion{Fe}{2} $\lambda$2757 line which originates above the ground state (1 eV) (c) $\phi =$ 10.404, and (d) $\phi =$ 13.641.  The dashed line represents 0 \kms, which is present from an IS cloud in \ion{Mn}{2} but not in \ion{Fe}{2}.  The sketch includes labeling of the projected Homunculus edge, located $\approx$1\farcs3 NNW of \ec\ and the ridge of backscattered starlight which is discussed in section \ref{sec:back}. Broad absorptions from multiple shells in the Homunculus, centered on $-$500 \kms, and the Little Homunculus, centered on $-$150 \kms, are readily apparent in the high-ionization state ($\phi =$~10.404) but  weaken and narrow by three cycles later ($\phi =$~13.641). Multiple spectro-images recorded across cycle 10 at these same PAs showed few changes in structure across the long high-ionization state. While both observations were accomplished at the same PA $=$ 332, the ensuing eighteen years between exposures led to a six-fold increase in stellar flux that further contributed an artificial flux from charge transfer effects (CTE) in the extended region below $-$0.7\arcsec\ at $\phi =$ 13.641 for b and d.}
\label{fig:long}
\end{figure*}

Nebular absorptions decreased greatly between the high-ionization states of cycles 10 and 13 in response to the large increase of FUV radiation (Figure \ref{fig:long}). On the spectrum of \ec, the broad absorptions centered on $-$500 \kms\ (Homunculus) narrowed to an apparently single absorption near $-$500 \kms\ while the weaker absorption centered on $-$150 \kms\ (Little Homunculus) nearly disappeared. The large decrease in absorption extended to the SSE (towards higher latitudes of the foreground Homunculus lobe), but remained broadened, though weaker beginning about 0\farcs3 NNW of \ec. The transition from narrow to broad absorption occurred beginning close to the projected position of \ec\ shifting towards lower velocities. Relating to the foreground lobe of the Homunculus, the absorptions dropped towards the cap while lesser changes occurred toward the projected edge of the lobe, about 1\arcsec\ to the NNW.

The measured stellar FUV radiation impinging on these structures increased nearly ten-fold between 2000 and 2018 which led to increased ionization \citep{Gull21a} while the measured stellar NUV radiation increased about six-fold. In contrast, the nebular NUV continuum changed marginally which is consistent with the radiation increase being due to the  dissipation of the occulter in LOS, not intrinsic brightening of \ec. Singly-ionized metals dominated during the high-ionization state of Cycle 10, but the increased FUV (and EUV) led to most metals becoming multiply-ionized in the high-ionization state of cycle 13. As a consequence, metal absorptions from these shells disappeared. 

Signatures of resonant and near-resonant transitions of multiply-ionized metals are deep in the UV, typically below Ly~$\alpha$. Hence signatures of shells with metals in higher ionization states became increasingly inaccessible in the \hst/STIS NUV and FUV spectral ranges. The complex, multiple shells  in the LOS from the  Homunculus and the Little Homunculus  virtually disappeared by $\phi =$ 13.641,  were nearly absent at the position of \ec\  but appeared to be less affected in the lobe walls. 

The noticeably  smaller change in absorptions to the NNW is not surprising given that the walls of the foreground lobe appear to be optically thicker. In part this is because the walls curve  towards the edge of the foreground lobe. Additionally, the butterfly boundary, a structure in the disk region located close to \ec,  extends from the NW across the projected position of \ec\ \citep{Chesneau05a}.

These longslit CCD spectra confirm that the decrease in absorption extends over significant angular and spatial distances across the foreground lobe of the Homunculus but do not provide information on how much more of the foreground lobe is affected. Nor do these spectro-images provide information about changes in the back lobes or the skirt region. Infrared and radio observations will be necessary to determine if additional regions of the Homunculus and interior structures are changing with time.

Changes of absorption across the long term for additional resonant and near-resonant lines are shown in Appendix \ref{sec:app}.
\subsection{Zooming in with STIS echelle spectro-images} \label{sec:echelle}

\begin{figure*}[ht]
\includegraphics[width=17.8cm]{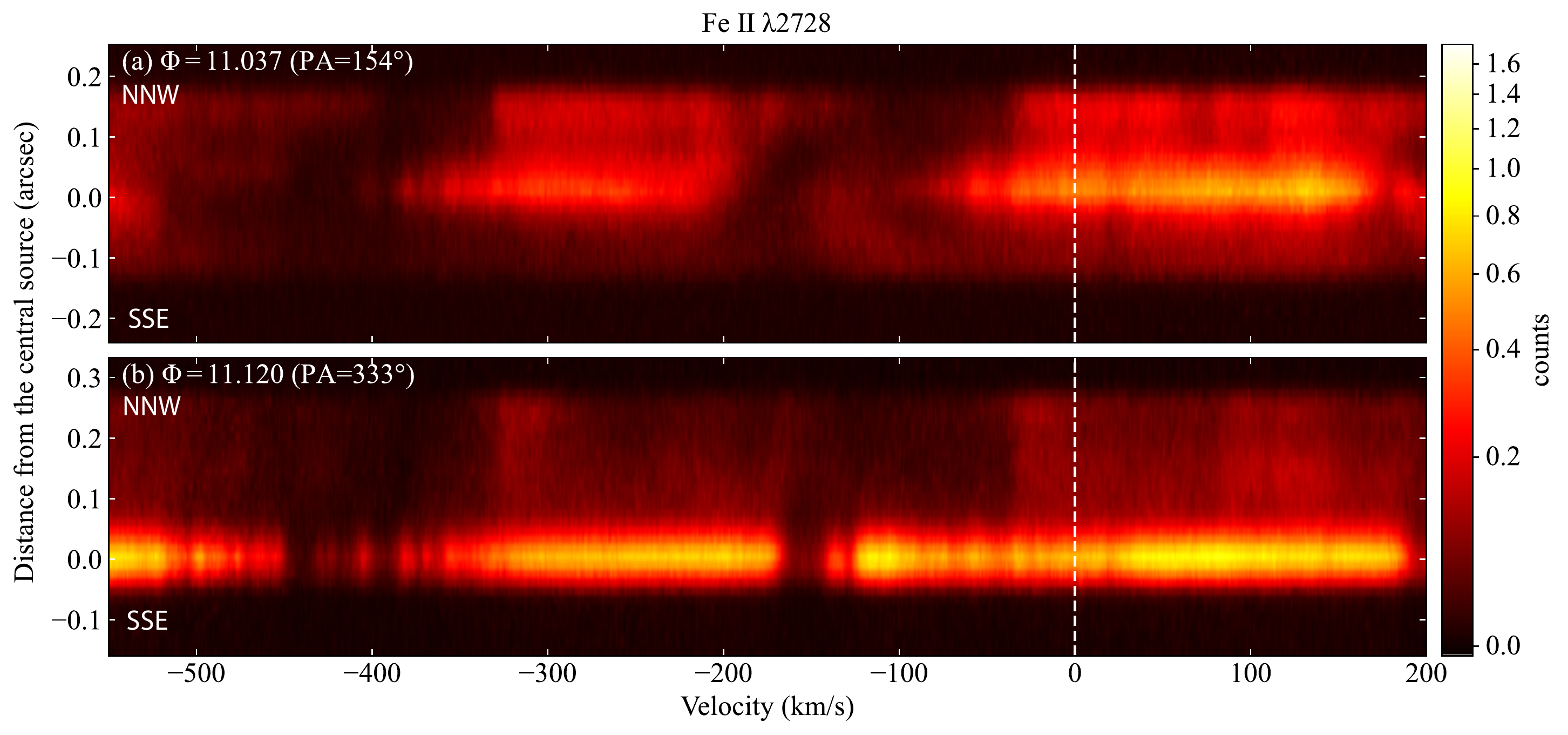}
\caption{Phase-dependent changes from low- to high-ionization state recorded in echelle spectro-images of  \ion{Fe}{2} $\lambda$2728 absorption structures:  
{\bf Top:} \ec\ was centered in the 0\farcs3$\times$0\farcs2 aperture deep in the low-ionization state ($\phi =$ 11.037).  Broad, high-velocity absorption from the interacting winds shifts from $\approx-$450 \kms\ SSE ($-$0\farcs15) to $-$550 \kms\ centered upon \ec\ and back to $-$450 \kms\ NNW ($+$0\farcs15), indicating that the Fe$^+$ wind structure is resolved and extends at least a few hundred au beyond the continuum core.  A narrow absorption, centered at $-$146 \kms\ on the star broadens in the NNW as   a continuous, shallow absorption extending to $-$35 \kms\ terminated by an emission feature originating from Weigelt B. 
{\bf Bottom:} A spectro-image recorded at $\phi =$  11.119, when \ec\ had returned to high-ionization state.  \ec\ was positioned at the edge of the aperture in an effort to map structures of Weigelt B and D and the intervening winds. The broad, shallow absorption extends NNW in front of Weigelt D.} \label{fig:occult}
\end{figure*}
The \hst/STIS achieves its best angular resolution in the NUV,  FWHM $=$ 0\farcs075, measured directly from the echelle spectro-images presented in Figure \ref{fig:occult}. The  0\farcs3$\times$0\farcs2 aperture, in combination with the E230H grating formats, resolves a portion of the winds and nebular components at a spectral resolving power, R~$=$ 110,000, for the stellar spectrum.   The nebular structure fills the full 0\farcs2 width of the 0\farcs3$\times$0\farcs2 aperture, and this leads to a lower nebular resolving power, R$\approx$35,000. Hence, the various absorption components are less-well resolved in the nebular portions of the spectro-image than on the stellar spectrum.

Despite the relatively small  aperture, the spectro-image  fits well within the spacing of NUV echelle orders on the format of the NUV MAMA allowing for proper detector background measures. This aperture provides some spatial information on changes in velocity structure close to the central star at high spectral resolving power.

Multiple narrow absorption lines are seen in the NUV echelle spectro-images of \ec. Figure \ref{fig:occult} displays the spatially- and spectrally-resolved velocity profiles of \ion{Fe}{2} $\lambda$2728 from two echelle spectro-images, one recorded during the low-ionization  and the other during the early high-ionization states of Cycle 11. Superimposed upon broad wind-absorption components are many narrow components, originating from the multiple velocity-differentiated shells in LOS previously catalogued by \cite{Gull06}. The broad absorptions, associated with colliding wind structures, extend beyond the continuum core of \ec\ ($\approx$ 0\farcs075), the apparent size of which is limited by the angular resolution of \hst, most likely  which is not the actual size of the continuum core in the NUV\footnote{\cite{Weigelt21} used the Very Large Telescope Interferometer to resolve the continuum structure in the mid-infrared to be 0\farcs0028 fwhm and imaged wind structures in the light of \ion{H}{1} Br$\alpha$ that extend $\approx$0\farcs026  around \ec. \cmfgen\ models indicate that the NUV continuum structure of \ec-A is likely more extended than in the infrared.}. 

The spectro-image, displayed in Figure \ref{fig:occult} Top, was recorded deep in the low-ionization state, $\phi =$ 11.037, within a day of the lower resolution spectro-images displayed in Figures \ref{fig:hilo}b, d. The broad absorptions, associated with the interacting winds, resolved from the continuum core of \ec, extend at least 0\farcs15 beyond on both sides, shifting from approximately $-$450 \kms\ from the SSE to $-$550 \kms\ on the star back to $-$450 \kms\ to the NNW. Multiple narrow absorption features are superimposed upon the point-like stellar continuum, but, due to the 0\farcs2 width of the 0\farcs3$\times$0\farcs2 aperture,  are  broadened as the nebular continuum fills the aperture height (0\farcs2) limiting the resolving power, R, to 35,000.

The spectro-image, displayed in Figure \ref{fig:occult}, Bottom, was recorded in the early high-ionization state, $\phi =$ 11.119 (within one day of the lower resolution spectro-images displayed in Figures \ref{fig:long}a, c)  with \ec\ placed at the edge of the aperture in order to record the spectrum of Weigelt D located to the NNW of the star. The broad interacting wind absorption is much weaker, permitting improved visibility of the multiple narrow, nebular absorption features in the LOS from \ec.

\subsubsection {High velocity absorptions seen in the echelle spectro-images}

Comparisons of the two spectro-images affirm changes of the stellar and nebular spectra between the low- and high-ionization states. The multiple, narrow-velocity shell absorptions, visible on the high-ionization stellar spectrum, are lost in the very strong, broad wind absorption during the low-ionization phase. Both narrow absorption complexes centered on $-$500 and $-$150 \kms\ are affected. 
\begin{figure*}[ht]
\includegraphics[width=16.5cm]{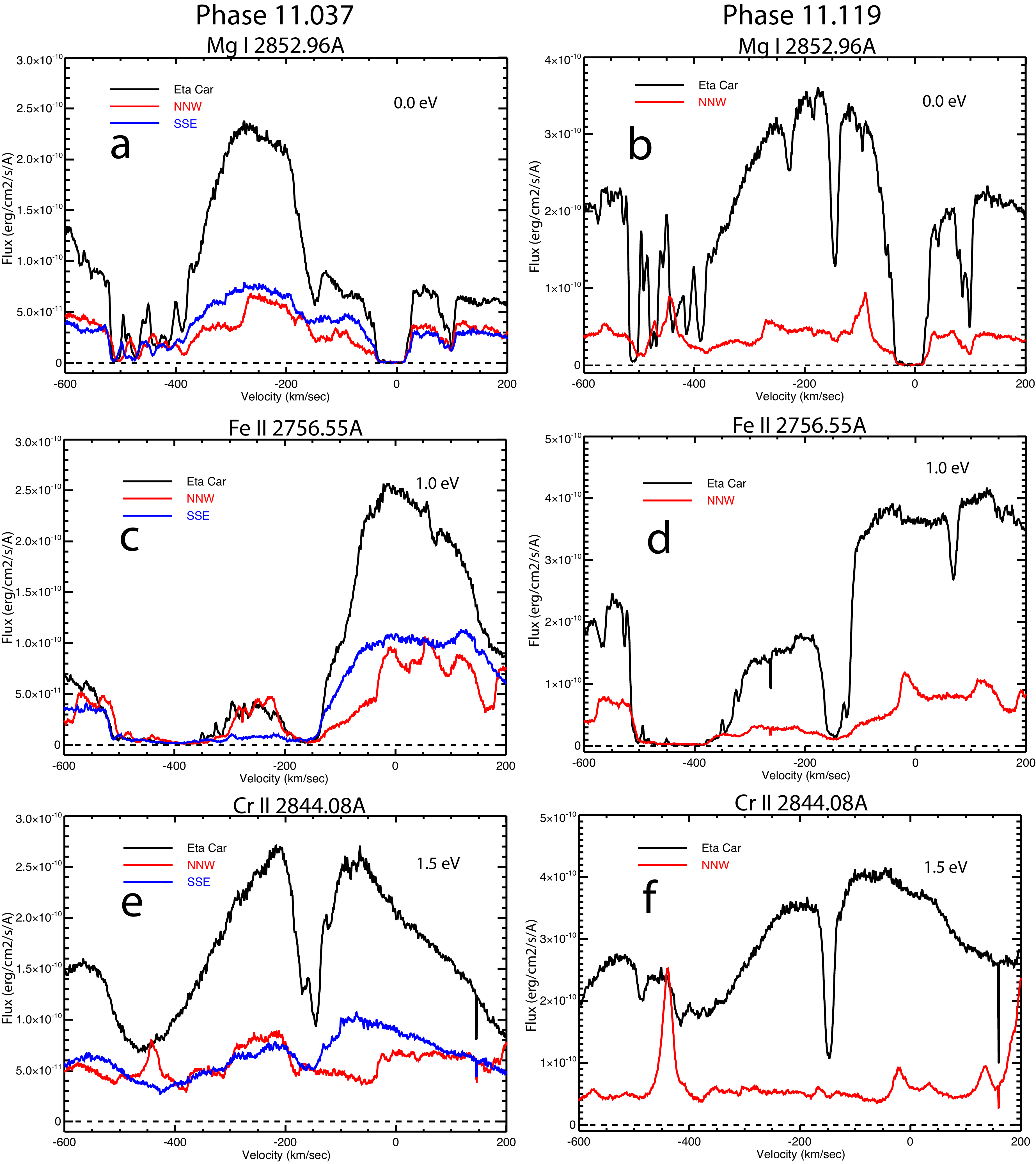}
\caption{Velocity profiles of three absorption lines that originate from significantly different energy levels during the low-ionization state, $\phi =$ 11.037 (left column) and the early high-ionization state $\phi =$ 11.119 (right column).  The \ion{Mg}{1} $\lambda$2853 line (0.0 eV) (top row) traces changes in  multiple shells where neutral gas existed. The \ion{Fe}{2} $\lambda$2757 line (1.0 eV) (middle row) is saturated at most shell velocities as singly-ionized metals dominate across the multiple shells. The \ion{Cr}{2} $\lambda$2844 line (1.5 eV) (bottom row) is dominated by the P~Cygni wind profile with strong absorption by shells with velocities between $-$120 and $-$190 \kms, but with only a few high velocity shells absorbing in the high-ionization state. The velocity profiles are (1) Eta Car, on the star, (2) {\color{red} NNW} which includes portions of the emission structure, Weigelt D, plus modified scattered starlight to the NNW and (3) {\color{blue} SSE} which is modifled scattered starlight. No nebular spectrum SSE of \ec\ was recorded in the echelle spectro-image recorded at $\phi =$ 11.119. Note: The
vertical scale on the left and right plots are different -- during periastron passage the NUV flux decreases, most likely a consequence of increased line blanketing by the primary wind.}  
\label{fig:eV}
\end{figure*}

More insight on the excitation changes between high and low-ionization states comes from examining three absorption line profiles (Figure \ref{fig:eV})  that rise from different energy levels: \ion{Mg}{1} $\lambda$2853 (0.0 eV) , \ion{Fe}{2} $\lambda$2757 (1.0 eV) and \ion{Cr}{2} $\lambda$2844 (1.5 eV). These velocity profiles were extracted from the same \hst/STIS echelle spectra that provided the spectro-images in Figure \ref{fig:occult}.

The extracted velocity profiles are:
\begin{itemize}
\item 
The spectrum of \ec\ extracted with a 0\farcs075-wide synthetic aperture for both echelle spectro-images.
\item 
The nebular spectrum, labeled SSE, was extracted with a 0\farcs100-wide synthetic aperture from the echelle spectro-image recorded at $\phi~=$~11.037, offset 0\farcs100 SSE from \ec, centered in the 0\farcs3$\times$0\farcs2 aperture. No nebular spectrum to the SSE was recorded at $\phi =$ 11.119.
\item 
The nebular spectrum, labeled NNW was extracted differently from the two spectro-images. 
\begin{itemize}
    \item For the spectro-image recorded at $\phi =$ 11.037, a 0\farcs100-wide synthetic aperture, offset 0\farcs100 NNW from \ec, centered in the STIS 0\farcs3$\times$0\farcs2 aperture. This enabled measurement of the nebular spectrum both  NNW and SSE of \ec\ at $\phi =$11.037. 
    \item For the spectro-image recorded at $\phi$~$ =$~11.119, a 0\farcs150-wide synthetic aperture  offset 0\farcs200 NNW as \ec\ had been placed close to the SSE edge of the STIS 0\farcs2$\times$0\farcs3 aperture.
\end{itemize}
\end{itemize}

Velocity profiles are dominated by modified P~Cygni wind-line contributions. \cite{Groh12}, using \cmfgen-based models of the binary winds, estimated the terminal velocity of \ec-A, \Vinf{$_{,A}$} $= -$420 \kms. The wind of \ec-B accelerates the wind-wind collision structures to higher velocities as is noticeable in velocity profile of \ion{Cr}{2} $\lambda2844$ in the low-ionization state (Figure \ref{fig:eV}e) compared to the high-ionization state (Figure \ref{fig:eV}f). The observed \Vinf\ shifts from $-$420 to $-$550 \kms\ \citep{Gull22}.

The \ion{Fe}{2} $\lambda$2757 line profile (Figures \ref{fig:eV}c, d) is saturated  across most of the range from $-$120 to $-$500 \kms\ but the \ion{Mg}{1} $\lambda$2853 profile in the high-ionization state (Figure \ref{fig:eV}b), allowing for the many narrow velocity components, has the general P~Cygni shape. In the low-ionization state (Figure \ref{fig:eV}a), the apparent P~Cygni profile appears to exceed $-$500 \kms\ but may be influenced by weaker shells. 

Narrow \ion{Mg}{1} $\lambda$2853 absorption structures are readily apparent against the stellar spectrum in both low- and high-ionization (Figure \ref{fig:eV}a,b). No high-velocity, narrow absorption is obvious in the \ion{Cr}{2} $\lambda$2844  line during the low-ionization state but does appear at $-$512 and $-$420 \kms\ during the high-ionization state (Figure \ref{fig:eV}c, d). Such is consistent with the high-velocity shells being partially ionized in both ionization states.

Strong absorption components are present between $-$100 and $-$200 \kms\ in all three lines. \ion{Mg}{1}, \ion{Fe}{2}, and \ion{Cr}{2}, for both ionization states indicating that the thermal excitation is significant in shells closer to the FUV source, \ec. Always present is the $-$146 \kms\ component which is associated with the Little Homunculus. In the high-ionization state,  the \ion{Fe}{2} $\lambda$2767 line has a strong absorption at $-$121 \kms\ and is saturated from $-$140 to $-$200 \kms\ in the low-ionization state. The \ion{Cr}{2} $\lambda2844$ line has a strong $-$146 \kms\ component in the high-ionization state, but adds two components, $-$121 and $-$168 \kms\ during the low-ionization state.

The nebular continuum in the low-ionization state, extracted 0\farcs1 SSE of \ec\ (labelled SSE in Figure \ref{fig:eV}), reflects the stellar continuum with saturated  absorptions in the \ion{Fe}{2} $\lambda$2757 line from $-$350 to $-$500 \kms\ and relatively narrow absorptions in \ion{Mg}{1} $\lambda$2853 superimposed upon the P~Cygni profile. 

In contrast, the nebular continuum in the low-ionization state, extracted NNW of \ec\ (labelled 'NNW' in Figure~\ref{fig:eV}) contains strong, broad absorptions when compared to the more-highly resolved stellar spectrum. Nebular line emission components originate from the background Weigelt D clump, plus broad absorptions extend from $-$35  to $-$150 \kms. The \ion{Mg}{1} $\lambda$2853 line includes a broad absorption extending from $-$270 to $-$370 \kms, then blends into narrow line absorptions extending to $-$510 \kms. The nebular continuum in the high-ionization state (recorded only to the NNW) again contains the very broad absorptions extending from $-$30 to $-$150 \kms.
\section{Where is the occulter/absorber located?}
\label{sec:LOS}
In section \ref{sec:longslit}, we demonstrated that the high velocity shells extend across significant areas of the Homunculus and Little Homunculus. The FUV absorber must have been much closer to the binary but extended well beyond our LOS. We turn now to examining spatial structures very close to \ec\ plus low velocity absorptions that provide clues to where the absorber appears to be located. In section \ref{sec:shell} we examine the spatial extent of recently-formed shells seen in forbidden emission. We then demonstrate apparent  expansion of the structure which backscatters NUV radiation in our direction (section \ref{sec:back}). Low velocity absorption components seen in atomic species in the FUV and in molecular species in the radio are discussed in section \ref{sec:loabs}. A sketch of the apparent geometry of the absorber relative to the recently formed shells is provided in section \ref{sec:geoclose}.

\subsection{Recently formed compressed shells by the interacting winds}\label{sec:shell}

\begin{figure}[ht]
\includegraphics[width=8.2cm]{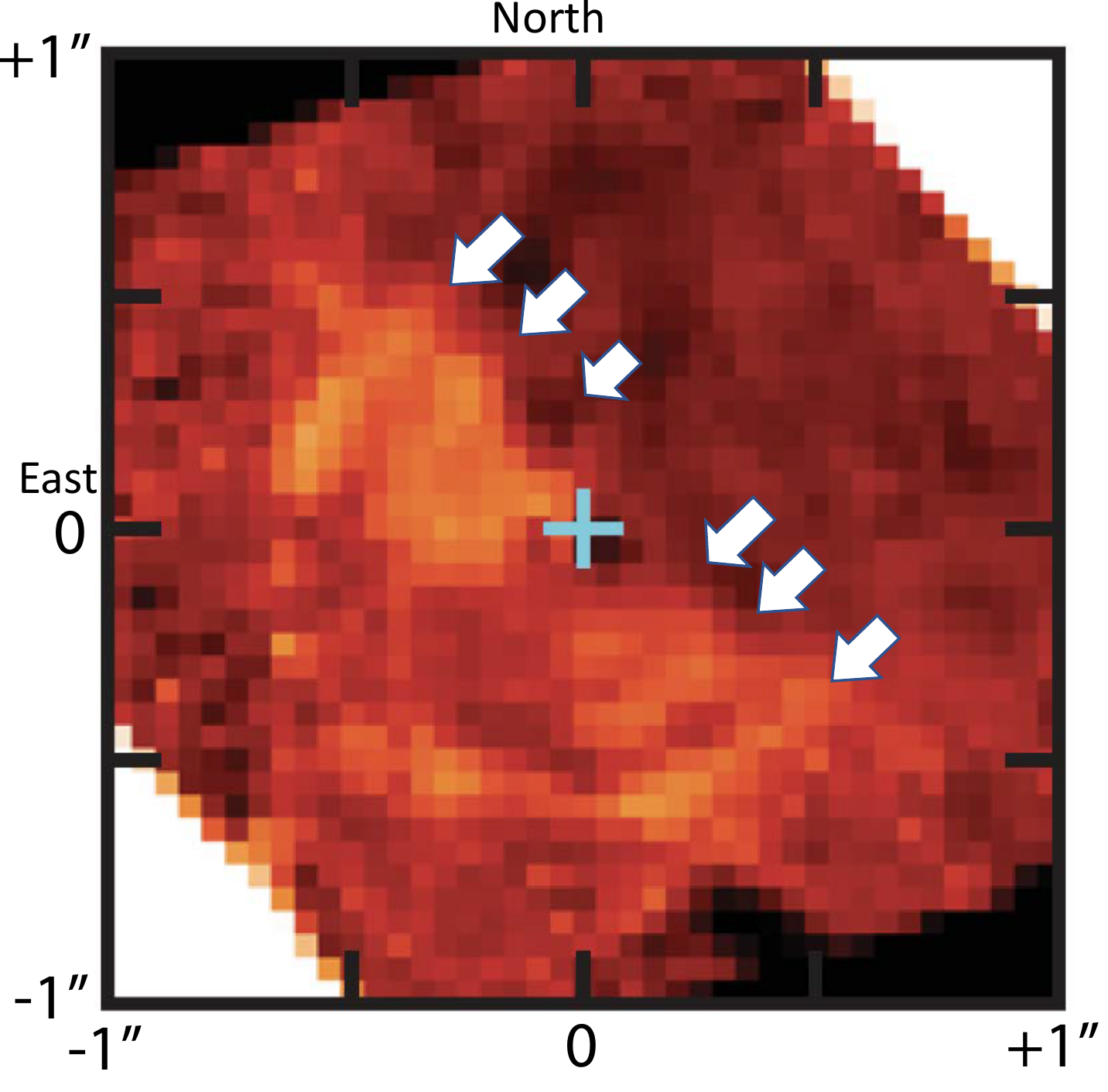}
    \caption{[\ion{Fe}{2}] shell expansion. The displayed image is a ratio of the [\ion{Fe}{2}] $\lambda$4815 line recorded at $\phi =$ 12.516 and 13.004 to enhance the expanding shells caused by passage of \ec-B through the extended continuous wind of \ec-A. The velocity range is 40 \kms\ centered on $-$60 \kms. The three pairs of arrows point to the ends of three shells found to be expanding outward. These shells do not exist to the northwest as the secondary wind has contiuously blocked the primary wind throughout the high-ionization state and the FUV from \ec-B has  ionized iron to higher ionization states (see Figure 6 in \cite{Gull16}\ for additional velocity frames). The $+$ symbol indicates the position of \ec.} 
    \label{fig:FeIImap}
\end{figure}

Three emission-line shells were identified in  [\ion{Fe}{2}] $\lambda$4815  and [\ion{Ni}{2}] $\lambda$7413  with \hst/STIS long-slit mappings of the central 2\arcsec$\times$2\arcsec\ recorded between 2010 and 2012 \citep{Teodoro13}. Three-dimensional cubes (RA, Dec and velocity) had been fabricated from  spectro-images recorded with the \hst/STIS 52$\times$0\farcs1 aperture positioned at 0\farcs05 spacing.  These cubes enabled two-dimensional, iso-velocity  images in forbidden emission lines.  The geometry of these shells formed nested hemispherical structures extending to the SE of \ec. These shells had formed from the dense, slow-moving primary wind compressed by the fast-moving secondary wind across each of three separate periastron passages. Accelerated from 420 to 475 \kms, the compressed shells persist as their  thermal velocities are much less than their expansion velocities. The projected distances, 400, 900 and 1400 au, as measured at $\phi =$12.680, are consistent with their origins occurring during the three periastron passages at the beginnings of cycles 10 through 12.

\cite{Gull16} extended the mappings in [\ion{Fe}{2}] $\lambda$4815 from $\phi =$~12.084 through 13.113 (2009 to 2015) to track the expansion of these shells across one orbital cycle. Figure \ref{fig:FeIImap}, a velocity slice from $-$40 to $-$80 \kms,  shows the difference in contrast by ratioing shell peaks to shell troughs and thus emphasizes the change in position  across one-half binary orbital cycle. This image demonstrates that, across cycle 12, the modulated primary wind extended over the hemispherical volume to the SE, but not to the NW.  The faster, less-dense secondary wind, over each lengthy, high-ionization state when \ec-B was beyond the extended wind of \ec-A,  evacuates the NW region with the remaining gas  being  highly ionized  by the FUV radiation from \ec-B. The slower, more-massive primary wind had been organized into shells moving to the SE by the secondary wind during each of the three previous periastron passages.

\subsection{The expanding dust structure behind the binary}\label{sec:back}
A noticeable 1\arcsec\ diameter structure, illuminated by scattered starlight,  is centered on \ec\ (Figures \ref{fig:hilo}, \ref{fig:long}). The background scattered starlight drops off at greater distances, but still records absorption from extended structures within  the Homunculus and Little Homunculus. The spatially-limited echelle spectro-images, with higher spatial and spectral resolutions (Figure \ref{fig:occult}), recorded variations of the extended wind of \ec-A. Such is  possible only if the scattered starlight originates from dust that lies beyond \ec\ and is back-, not forward-, scattered. The dusty structure that scatters the starlight lies beyond \ec\ and likely is very slow-moving ejecta originating with  one or both of the historical events.

\begin{figure*}
\includegraphics[width=17.cm]{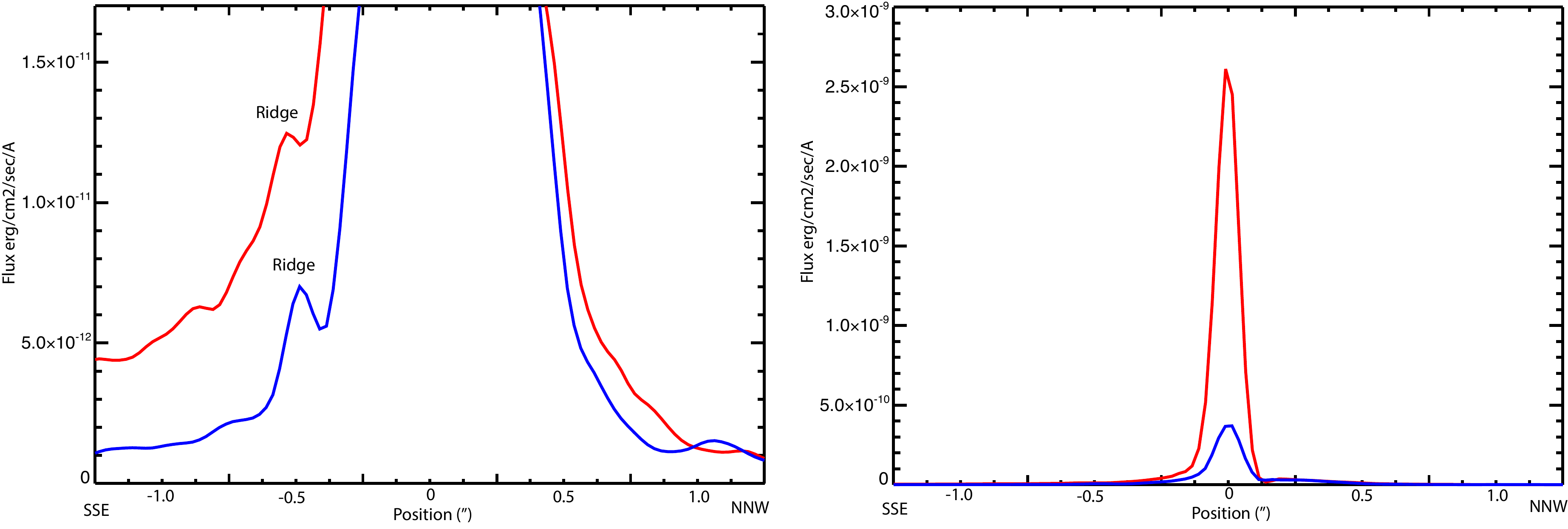}
\caption{Comparison of the spatial profiles for spectroimages recorded at $\phi =$ {\color {blue}10.404} and $\phi =$ {\color{red}13.641} ((see Figure \ref{fig:long}).   Left: Cross section showing the shifts of the SSE ridge at $-$0\farcs5 and the NNW edge at $+$0\farcs5. Expansion is notable of the ridge and valley to SSE and possibly the edge to the NNW. Right: \ec\ profiles. While the stellar flux in the NUV increased six-fold between 2000 and 2018, the measured FWHM was 0\farcs11 for both stellar profiles.  Note that both plots are in flux units, not normalized.}\label{fig:expand}
\end{figure*}

This scatterer appears to be expanding, most notably to the SSE.
A well-defined ridge of scattered-starlight, located 0\farcs5 SSE of \ec\ in Figure \ref{fig:long}, shifted outward between 2000 ($\phi =$ 10.404) and 2018 ($\phi =$ 13.641), when the spectro-images were recorded.  Spatial slices of the spectral images are compared in Figure \ref{fig:expand}. Measurements of the SSE ridge position, done with four spectro-images extending from 2540 to 2850\AA\ and 3900 to 3950\AA, showed this ridge   shifted outward 0\farcs063 over the eighteen year interval. The edge of the scattered-light structure to the NNW was also measured, yielding a shift of 0\farcs047 over the eighteen year time-interval. Adopting the accepted distance of \ec, 2300 pc, yields expansion velocities of 39 $\pm\ $3 \kms\ to the SSE and and 28 $\pm$\ 3 \kms\ to the NNW,  which is comparable to the radial velocity of the Weigelt clumps ($-$45 \kms, \cite{Zethson01}). Additional structures at 1 to 2\arcsec\ distance from \ec, associable with the Little Homunculus, are likewise seen to be moving outward, but are less well defined.

However, charge transfer inefficiency (CTI), caused by accumulating radiation damage to the CCD, artificially impacts the measured nebular expansions. The CTI effect increased over the eighteen year interval and the NUV brightness of \ec\ also increased. The ridge to the SSE is easily identified, as well as fainter structures in the Homunculus and the Little Homunculus that also expanded. The apparent expansion to the NNE of the scatterer is less well-defined. Hence the apparent expansion to the NNW must be considered an upper limit, not a true measure.  Indeed, the [\ion{Fe}{2}] shells as shown in Figure \ref{fig:FeIImap} strongly suggest that the secondary wind and ionizing flux of \ec-B may limit the expansion of the background dusty structure to the NW.

\subsection{Low velocities associated with the absorber}\label{sec:loabs}

The infrared flux of \ec, summarized by \cite{Mehner19}, has not changed substantially over the past half century, which is consistent with minimal changes of the binary properties over that period of time.  Stable wind properties over the  past fifty years should have produced more than the three shells seen in [\ion{Fe}{2}]  \citep{Teodoro13,Gull16} located at greater distances, analogous to the multiple, regularly-spaced shells recently detected in dust  emission and forbidden radiation surround WR140  \citep{Lau22}. While the ionizing radiation leading to Fe$^+$ may be depleted, singly ionized iron absorption is traceable within the Little Homunculus and Homunculus, well beyond these shells. Instead, the massive, slow-moving occulter/absorber has been continuously stopping and absorbing previous shells in the form of shocks.

Based upon the expansion velocity of the three shells, the extended structures associated with the occulter could be located just beyond the outermost shell seen by \cite{Teodoro13}, i.e. just beyond 1400 au radial distance from \ec. We therefore searched for velocity signatures in the spectrum of \ec\ that are consistent with a structure ejected in the 1840s or 1890s that  exceeds that distance.

The occulter in front of the binary could be a clump of similar nature as the Weigelt clumps, B, C and D \citep{Weigelt86} and additional clumps seen in the infrared Butterfly nebula \citep {Chesneau05a} and likewise clumps seen in [\ion{Fe}{2}] \citep{Gull16}. Adaptive optics imaging in  [\ion{Fe}{2}] $\lambda$1.644$\mu$ of multiple structures within projected distances of 4000 au from \ec\ provided epochal origins that ranged from the early 1830s to the late 1910s, being consistent with the Great and Lesser Eruptions \citep{Artigau11}. While current telescopes and instrumentation are unable to separate extended structure in front of the much brighter \ec, most likely the occulting structure was ejected in this time interval.

Such an occulter scenario seems likely because \cite{Zethson01} measured the  radial velocity of Weigelt B and D to be $-$45 $\pm$ 4 \kms. Similar low radial velocities have been measured in front of the binary at radio wavelengths and in the UV using \hst/STIS echelle spectra, as discussed in the next two paragraphs. 

\cite{Chesneau05a}  obtained NIR imagery of \ec\ and structures within the central region with 0\farcs1 angular resolution that showed structure around \ec\ extending about 0\farcs5 to the NW across the position of \ec\ and about 0\farcs2 to the SE (see their Figure 5). The   Weigelt clumps C and D, located 0\farcs1 to 0\farcs3 to the NW,  are directly  ionized by the FUV of \ec. They appear to be surface blisters of the dusty, extended  structure mapped by \cite{Chesneau05a} as noted by \cite{Teodoro19}. 

\cite{Bordiu19} mapped the central  region at 345.8 GHz with {\it ALMA} with 0\farcs2 angular resolution  that confirmed continuum emission extending from \ec\ to the NW (see their Figure 2). They detected line absorption in HCN and H$^{13}$CN against the continuum from \ec\ at  V$_{lsr} = -$60 \kms\  (V$_{helio} = -$49 \kms).

The central 2\arcsec$\times$2\arcsec\ region was mapped in [\ion{Fe}{2}] $\lambda$4816 with R$ = \lambda/\delta\lambda =$ 10,000 and  0\farcs1 angular resolution \citep{Gull16}. Their Figure 3, which shows changes across one binary orbital cycle, shows a bright emission structure to the NW that includes Weigelt C and D. Re-examination of this data demonstrated that this narrow-line emission structure extends to within 0\farcs1 of \ec, but is not present to the SE. The slowly varying central velocity terminates close to \ec\ at $-$49 $\pm$ 4 \kms, the velocity of the molecular absorptions published by \cite{Bordiu19}. \cite{Weigelt12} found the origin of the Weigelt clumps to be consistent with the 1890s event. 

 \cite{Smith04c} measured the system velocity of the Homunculus to be V$_{helio} = -$8.1 $\pm$ 1  \kms. \cite{Nielsen07b} adopted that velocity for the binary system velocity, finding consistency while deriving the binary orbital parameters. If we assume this to be the central velocity of the binary,  the expansion velocity, $-$41 \kms, originating from the 1890s event with no intervening material, translates to 1000 au.  For comparison, \cite{Teodoro13} found  the outermost shells to be 0\farcs8 (1850 au) distant from \ec, but also noted that the shells appeared to been moderately accelerated.
The system velocity of the binary  could be slightly different. We do not know if the bipolar ejection of the Homunculus was symmetric for both lobes. The expansion velocity of the occulter could be significantly (smaller) larger in the LOS. If the system velocity for the binary were as much  different than 10 \kms, the distance estimate  ould (decrease) increase to a (750 au) 1250 au. 

\begin{figure}
    \centering
    \includegraphics[width=8.5 cm]{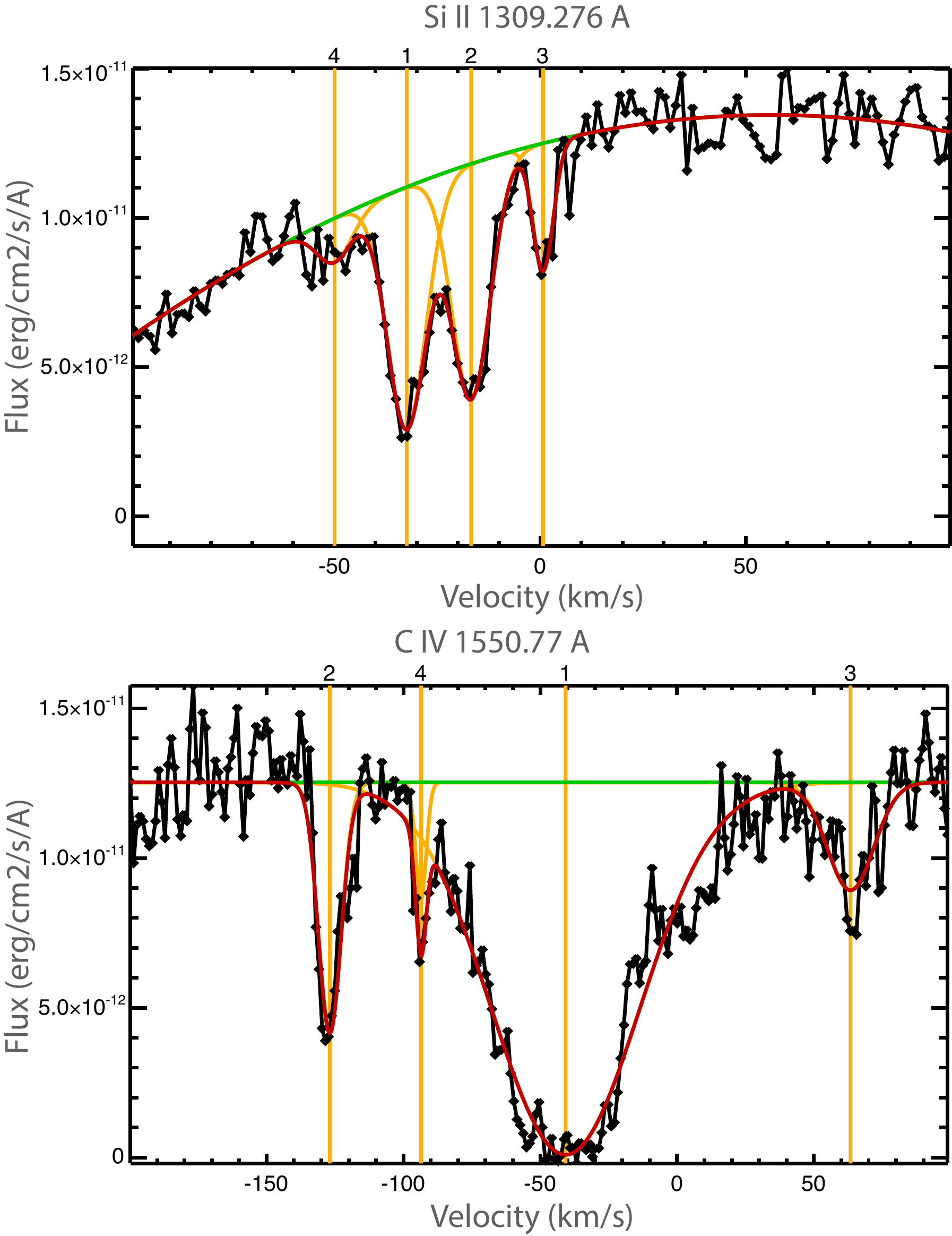}
    \caption{ Low velocity absorptions attributable to the occulter. {\bf Top:} \ion{Si}{2} $\lambda$1309 velocity profile. Two strong absorptions centered at $-$18 and $-$32 \kms\ with two weaker components at $-$50 and $+$1 kms (foreground cloud). {\bf Bottom:} \ion{C}{4} $\lambda$1551 velocity profile with a strong absorption entered at $-$39 \kms\ and extending from $-$20 to $-$70 \kms. Observation recorded at $\phi =$ 13.886.}
    \label{fig:vellos}
\end{figure}

Additional searches were done for similar absorption velocities in the \hst/STIS UV echelle spectra recorded in both high- and low-ionization states (Table \ref{tab:obs}). Two low-velocity absorption profiles were measured in \ion{Si}{2} $\lambda$1309 and \ion{C}{4} $\lambda$1551 (Figure \ref{fig:vellos}). The two strongest absorptions in \ion{Si}{2}, $-$32 and $-$18 \kms, presumably are from slower, singly-ionized structures, which is to be expected as many weaker analogues of the Weigelt clumps were found at similar velocities in the central 1\arcsec\ diameter region in [\ion{Fe}{2}] $\lambda$4815 \citep{Gull16}. A relatively weak absorption is  present at $-$50 \kms. 

A very strong velocity component is present in \ion{C}{4}~$\lambda$1551 centered at $-$40 \kms, extending from $-$100 to $+$20 \kms\ (Note that this \ion{C}{4} resonant line is the red member of the well-known doublet.  Absorption at this velocity from the shorter wavelength member, $\lambda$1548, is masked by the strong, wind-wind, blue-shifted component of $\lambda$1551). Both \ion{Si}{4} $\lambda\lambda$1304, 1403 velocity profiles (not shown) are completely saturated between $-$60 and $-$20 \kms\ which reinforces the presence of the $-$50 \kms\ gas somewhere in the LOS. The strengths of these absorption profiles changed little from high-ionization to low-ionization states across both periastron passages, 10/11 and 13/14, despite the considerable change in FUV flux. The strong \ion{C}{4} $\lambda$1549 is suggestive of a shock which could be evidence of ongoing shell collisions with the occulting structure. This low-velocity profile of \ion{C}{4} persisted across all FUV spectra recorded by the STIS across Cycles 10/11 and 13/14.

If mappings continued across cycles 13 and 14, we suggest that the outermost shell would have continuously disappeared, having collided with the occulter/absorber with a new shell forming close to \ec. However, with the disappearance of the occulter, would that be evidence that the absorber has completely dissipated? Only followup observations would tell. High spatial resolution studies of molecular and ionic species in the radio spectrum may provide additional insight along with future UV/visible studies.

We conclude that the occulter is  1000 au from  \ec\ with a lower limit of 500 au to 2000 au.

\subsection{The geometry of the absorber}\label{sec:geoclose}

\begin{figure}[ht]
\includegraphics[width=8.5cm]{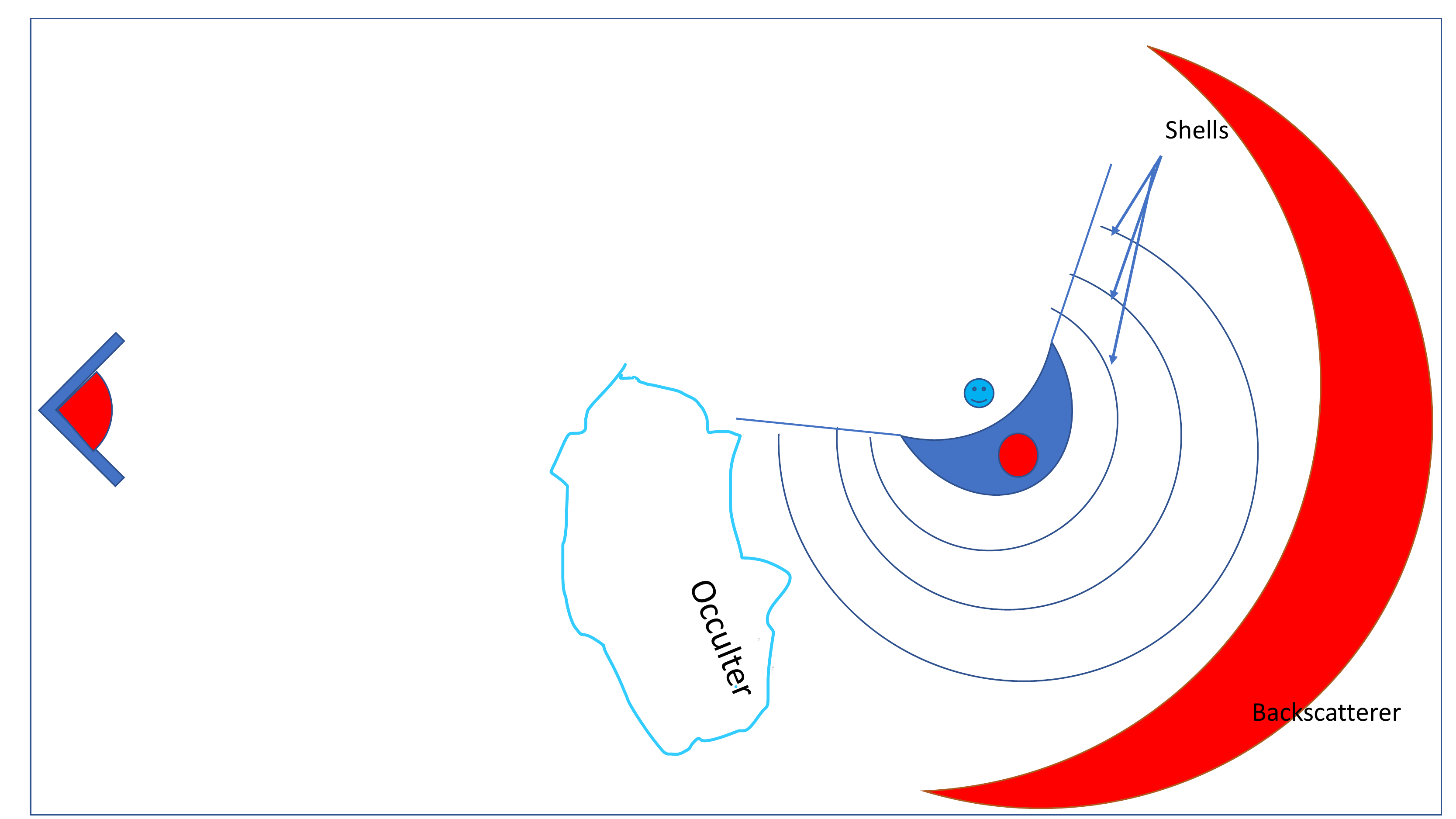}
\caption{Sketch of the structures in close proximity to the binary ({\bf NOT TO SCALE}).  The massive primary, {\color{red}\ec-A}, has a very extended wind structure ({\color{blue} dark blue}) which is modified by the lesser, but faster, wind of {\color{blue}\ec-B}. Three shells of compressed primary wind, seen in [\ion{Fe}{2}] $\lambda$4815 \citep{Teodoro13} are located beyond the current wind structure of the binary. The occulter is a structure that extends to one side of the binary. The occulter blocked one side of the extended primary wind and strongly blocked the central core, but with the occulter dissipation, our LOS to the core has become increasingly transparent.
On the far side of \ec\  a dusty structure scatters light from the binary back towards the observer.\label{fig:concept2}}
\end{figure}

A schematic of the extended occulter structures close to \ec\ is presented in Figure \ref{fig:concept2}.
 The three recently-formed shells, as mapped in 2012, lie closest to \ec\  with the occulter just external to the outermost shell. Under the scenario presented above, by 2022, two more shells would have formed while the two outermost shells have been destroyed by collision with, and absorption by, the occulter. At any given time  three shells  fill a volume to the SE somewhat larger than a hemisphere based upon the expansion imagery noted by \cite{Gull16} and the remaining hemisphere to the NW with  high ionization by the FUV flux from \ec-B. 

The occulter partially obscures our LOS from \ec. \cite{Hillier06} noted that the equivalent width of H$\alpha$\ in the direct stellar spectra, recorded in the early 2000s, was twice that predicted by \cmfgen\ models \citep{Hillier06}. They explained  that the absorber occulted the central core (where the continuum originates) more than it did of the primary wind. Indeed over the past two decades, the equivalent width of H$\alpha$ has decreased, trending towards the value seen in  scattered starlight from portions of the Homunculus \citep{Damineli21}. Direct imagery with the \hst/Faint Object Camera (FOC) in the UV \citep{Weigelt95} previously showed the nebulosity surrounding \ec\ to be clumpy in nature and also demonstrated the high relative brightness of the three ejected clumps. This additionally suggests the existence of an extended, nonuniform occulter.

The dusty structure that backscatters light from \ec\ is sketched beyond \ec. The periastron passage is in this general direction and suggests that a large amount of material may have been ejected during one or both of the two eruptions. Perhaps radio observations of molecular emissions may provide additional information on this structure. The broad absorption features to the NNW in Figures \ref{fig:hilo} through \ref{fig:long} are suggestive of interaction of the winds with this structure.

\subsection{Origin of the occulting structure} \label{origin}
The velocities of \ion{C}{4}, $-$40 \kms, is close to the velocity measured in HCN and H$^{13}$CN, $-$49 \kms, which suggests a related origin, namely the occulting structure. Moreover these velocities are close to the velocities measured of Weigelt BD, $-$45 \kms, by \cite{Zethson01}. \cite{Weigelt12} found the proper motion of  Weigelt D corresponded to an origin of 1880 $\pm$ 20 years, which is consistent with an origin from the Lesser Eruption of the 1890s.

The occulting structure must be well within both the Homunculus and the Little Homunculus as both structures responded to the long term changes as demonstrated in Figure \ref{fig:long}. Moreover the much slower-moving, occulting structure, were it ejected in the Great Eruption, would likely have blocked the multiple shells with velocities exceeding its 50 \kms\  velocity. Indeed the occulting structure has protected the multiple shells from the apparently ongoing,  interacting winds found to be expanding at 470 \kms\ \citep{Teodoro16}. Only  three recently-formed shells were seen in [\ion{Fe}{2}] and  [\ion{Ni}{2}]. Earlier shells appear to have been blocked and absorbed by the occulting structure.

An alternative explanation might be that the occulter was moving tangentially as a knife edge blocking \ec\ across the past half century. However it appears unlikely that such a structure would uncover \ec\ and the very extended region as seen in Figure \ref{fig:long} within the observed 18 year interval.  Occam's razor leads to the simplest explanation, namely a dissipating occulting structure.

\section{Conclusions}{\label{sec:SUM}}

 The brightening of \ec\  over the past two decades has been attributed to changing properties of the interacting winds. \cite{Davidson18} claimed that the mass-loss rate of the primary is decreasing. This contradicts the repeatability of the increase in the depth of the P Cygni absorption \citep{Damineli22}.  
\cite{Espinoza21a} pointed out that the X-ray flux measures at apastron have not changed appreciably nor has the X-ray light curve changed noticeably in the multiple periastron passage approaches that have been closely monitored. Recovery of the post-periastron flux towards high-ionization state does vary from passage to passage and is thought to be governed by details of the reforming bowshock structure as it is rebuilt from the clumpy LBV (primary)  wind  \citep{Davies08}. This suggests further investigation.

Figures \ref{fig:hilo}, \ref{fig:long}, \ref{fig:apcompare} and \ref{fig:ap2} demonstrate that a large portion of both the Homunculus and especially the Little Homunculus foreground lobes became more highly ionized between $\phi =$ 10.602 and 13.641. Indeed the velocity signatures of singly-ionized iron, that traced the structure of  the Little Homunculus  at $\phi =$10.404, completely disappeared by $\phi =$ 13.641 and the broad absorptions tracing the Homunculus extending from SSE of \ec\ to the position of \ec\ narrowed to the $-$512 \kms\ component. The singly-ionized metal lines are still present to a lesser degree in the nebular spectrum west of \ec\ at $\phi =$ 13.641 and  extend well beyond the Weigelt clump D. The decline in extinction by the occulter produced a ten-fold increase in FUV flux and higher nebular ionization with destruction of H$_2$ at $-$512 \kms\ in the intervening years from  2004 to 2018.

 This study demonstrates that the multiple shells within the Homunculus and Little Homunculus have changed in ionization over the past two decades. This is a direct result of the tenfold increase in FUV radiation noted by \cite{Gull21a,Gull22}. The studies by \cite{Damineli21} and those summarized therein show that the flux increase is caused by a dissipating occulter, first suggested by \cite{Hillier92a}. 

The long-term increase in the LOS P-Cygni absorptions may seem contradictory to the disappearance of nebular absorption in the UV. However, the strengths of the P-Cygni absorptions in the nebular-scattered starlight, were much stronger three decades ago. The explanation remains that dust in our LOS was blocking the continuum-emitting core more so than the much-more extended primary wind, which is the source of the P-Cygni profile. As noted by \cite{Mehner19}, all evidence points to a stable flux from the massive binary over the past half-century. The apparent increase in FUV flux, due to the dissipating extinction in our LOS, led to the ten-fold increase in the FUV and resulted in higher ionization of nebular shells in the foreground Homunculus lobe and destruction of H$_2$.

The combination of \hst/STIS spectro-imagery with published studies in the visible \citep{Weigelt86}, in the infrared \citep{Chesneau05a} and the radio \citep{Bordiu19} leads to a picture that the occulter in front of the binary is an extended clump of similar nature to the Weigelt clumps B, C and D and fainter clumps seen in the infrared Butterfly nebula.

Destruction of the intervening dust from the occulter in the long term has increased ionization of multiple shells within the Homunculus from neutral and singly-ionized metals to multiply-ionized metals. Resonant lines for multiply-ionized metals are in the EUV. More FUV radiation now reaches throughout the Homunculus. Eventually, as the Homunculus continues to expand,  most dust within the Homunculus is likely to be modified and/or destroyed. With dust destruction, the scattered light that currently traces the spatial structure of the Homunculus, will gradually disappear. The ejecta, as it becomes increasingly diluted, will mix with the interstellar medium, currently at much larger distances due to previous winds from \ec\ driving out the ISM.

The Homunculus, when the dust is destroyed and the associated gases multiply-ionized along with continued shell expansion, will become nearly transparent as likely has happened with ejecta from the winds of other massive binaries. \ec\ and its ejecta, already known as an astrophysical laboratory, as future observations are made, will continue to provide insight into the evolution of ejecta from massive stars.

This study, limited to line absorptions against \ec\ and nebular-scattered starlight, has focused on changes in LOS from \ec\ and a relatively small area surrounding \ec. Changes are obviously occurring throughout the Homunculus as the UV radiation eats away at the dust and molecules along with the ongoing expansion.

Future studies, especially in the FUV, become all the more important. Hopefully the \hst/STIS access will continue through the next decade as monitoring of changes across the next periastron in 2025 now become key to confirming the changes seen across periastron passage 14. Important studies of the entire Homunculus are now warranted in the infrared and radio spectral regions. Specifically, radio studies of atomic and molecular shocks in the close vicinity of \ec\ should be seriously considered.
And along with ongoing studies,  modeling of the massive binary system and its wind interactions and its radiation on the expanding ejecta continue to be needed.

\begin{acknowledgments} NR acknowledges funding from \hst\ programs 15611 and 15992 which were accepted as supplementary observations associated with CHANDRA programs 20200564 and 21200197. MFC is supported under the CRESST-II cooperative agreement \#80GSFC17M0002 with the NASA/Goddard Space Flight Center. AD received funding from FAPESP 2011/51680-6 and and CNPq 301490/2019-8. AFJM is grateful for financial aid from NSERC (Canada). 
\end{acknowledgments}
\facilities{HST/STIS}
\bibliography{ref}{}
\bibliographystyle{aasjournal}
\appendix{}
\renewcommand\thefigure{A.\arabic{figure}}
\setcounter{figure}{0}
\renewcommand\thetable{A.\arabic{table}}
\setcounter{table}{0}
\section{Nebular absorptions 2560-2610\AA}\label{sec:app}
\begin{figure*}[ht]
\includegraphics[height=6.4cm]{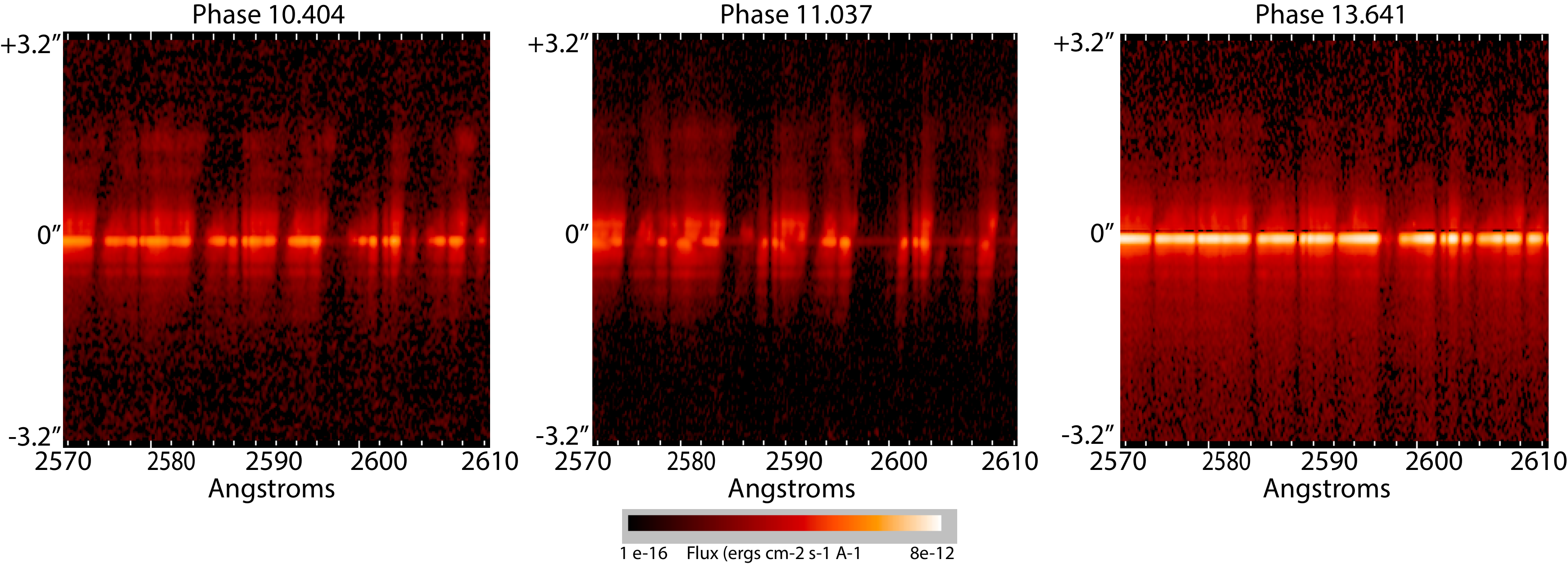}
\caption{Changes in absorption lines originating from the ground state.  Within one orbital cycle strong absorptions in  the high-ionization state (left: $\phi =$ 10.404) increase to even stronger absorptions in the low-ionization state (middle: $\phi =$ 11.037). Two broad velocity complexes  centered on $-$500 \kms and $-$150 \kms\ at the star position extend from $\approx -$550 and $-$200 \kms\ SSE (below the position of \ec) to near 0 \kms to the NNW (above the position of \ec). Three orbital cycles later, the absorptions in the high-ionization state have all but disappeared in the SSE  (right, $\phi =$ 13.641) caused by the ten-fold increase in FUV radiation. Only one component, $-$513 \kms\ associated with the Homunculus, remains SSE (bottom) of \ec, but a broader component extending from the star position remains  to the NNW which demonstrates that the absorption remains in the walls at lower elevations of the Homunculus  foreground lobe. 
The 0 \kms\ IS absorption lines serve as velocity reference for each of the shell structures.  The leftmost absorption line, \ion{Mn}{2} $\lambda$2577, is described in more detail and in velocity space in Figures \ref{fig:hilo} and \ref{fig:long}.
\label{fig:apcompare}}
\end{figure*}

\begin{figure*}
\includegraphics[height=6.4cm]{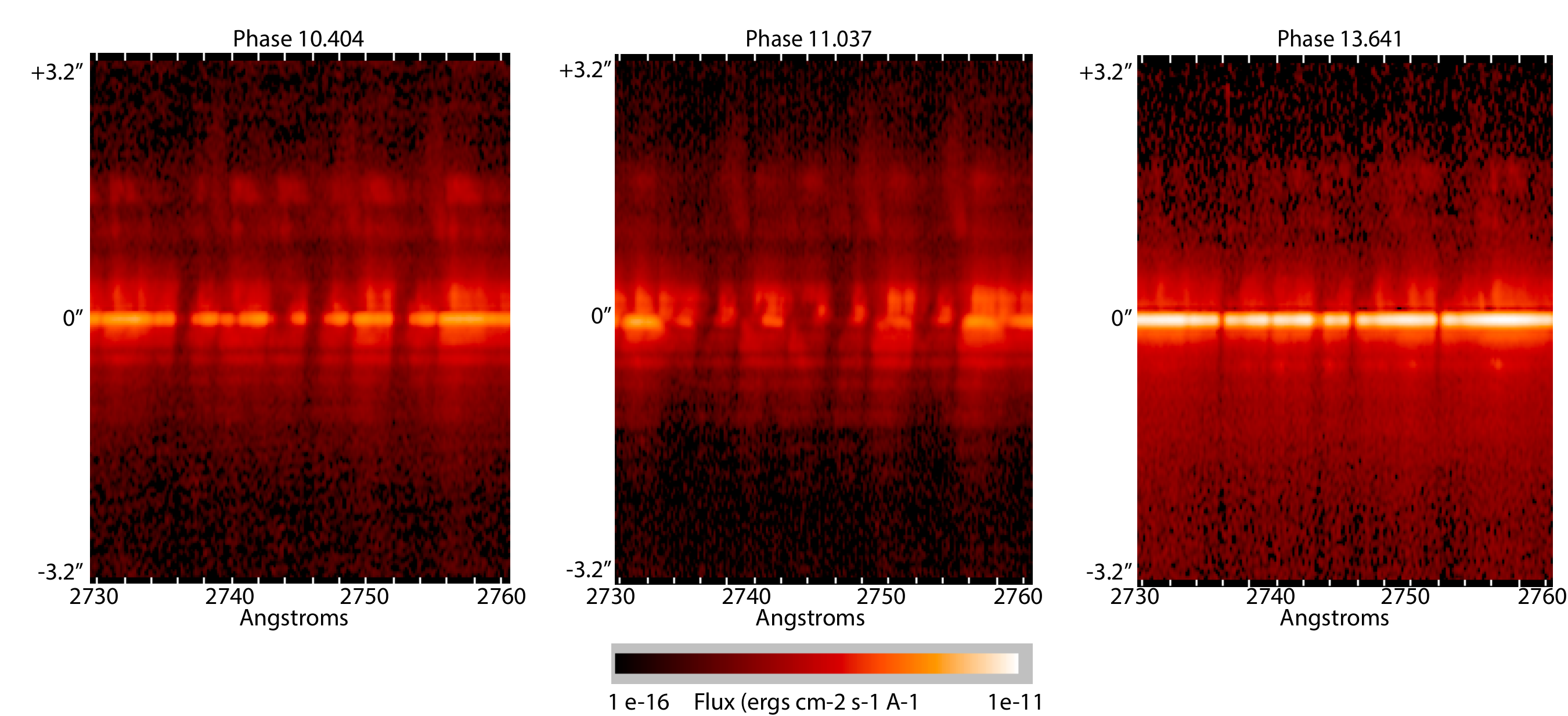}
\caption{Comparison of changes of \ion{Fe}{2} absorption lines originating from the 1.0 eV energy level. As with the ground state absorbers, the absorption increases between the high-ionization state (left, $\phi =$ 10.404) and the low-ionization state (middle, $\phi =$ 11.037). The absorptions nearly disappear in the SSE (below the position of \ec), other than the $-$ 513 \kms\ component, in the high-ionization state three orbital cycles later (right, $\phi =$ 13.641). But likewise the broad absorption remains to the NNW (above the position of \ec) at lower latitudes of the foreground lobe. The right-most absorption line, \ion{Fe}{2} $\lambda$2757, is presented in more detail and in velocity space in Figures \ref{fig:hilo}  and \ref{fig:long}, 
\label{fig:ap2}}
\end{figure*}

Figures \ref{fig:apcompare} and \ref{fig:ap2}  demonstrate that the absorption profiles shown in Figure \ref{fig:hilo} occur repeatedly in multiple metal lines in the NUV with details influenced by the physical conditions and the transition probablities of each metal line. In simplest terms, the ionization state of the shells within the Homunculus are strongly influenced by ionizing FUV radiation. The absorptions from singly-ionized metals increase  across the binary orbit from high-ionization ($\phi =$ 10.404) to low-ionization ($\phi =$ 11.037) states. The absorptions in the two high-ionization states ($\phi =$ 10.404 and 13.641) should be very similar. However the increased FUV radiation has increased the ionization levels of the shells; the trend of disappearing absorption is general. The strongest lines are listed in Table \ref{tab:alines}.

The plots are displayed with the y-axes  the same in angular interval for direct comparison, but the x-axes span different spectral ranges. Many of the \ion{Fe}{2} lines extend beyond the continuum background, which terminates at about 1\farcs8 NNW, changing from absorption to emission, characteristic of the oft-described P~Cygni profile, but in this case is caused by the spatially-resolved nebular structure.

\begin{table}
\tablenum{A.1}
    \centering
    \caption{Identified lines of interest} 
    \label{tab:alines}
   Figure \ref{fig:apcompare} lines\\
    \begin{tabular}{lrcr}
Ion & Wavelength & Energy levels& Lower Energy\\
       \hline
Mn II &2576.88 & 4s a$^7$S$_3$ -- 4p z$^7$P$_4$& 0.0 eV\\
Fe II &2586.65& 4s a$^6$D$_{9/2}$ -- 4p z$^6$D$_{7/2}$ & 0.0 eV\\
Fe II &2600.17 & 4s a$^6$D$_{9/2}$ -- 4p z$^6$D$_{9/2}$ & 0.0 eV\\
Mn II &2594.50 & 4s a$^7$S$_3$ -- 4p z$^7$P$_3$ & 0.0 eV\\
Mn II &2606.46 & 4s a$^7$S$_3$ -- 4p z$^7$P$_2$ & 0.0 eV\\
\hline\\
\end{tabular}  
\\
 Figure \ref{fig:ap2} lines\\
    \begin{tabular}{lrcr}
Ion & Wavelength & Energy levels& Lower Energy\\
       \hline
Fe II &2740.36 & 4s a$^4$D$_{7/2}$ -- 4p z$^4$D$_{7/2}$ & 1.0 eV\\
Fe II &2747.30 & 4s a$^4$D$_{3/2}$ -- 4p z$^4$F$_{5/2}$ & 1.1 eV\\
Fe II &2750.13 & 4s a$^4$D$_{5/2}$ -- 4p z$^4$F$_{7/2}$ & 1.1 eV\\
Fe II &2756.55 & 4s a$^4$D$_{7/2}$ -- 4p z$^4$F$_{9/2}$ & 1.0 eV\\
\hline\\
\end{tabular}
\\
\end{table}

\end{document}